\documentclass[aps,pra,twocolumn,showpacs,superscriptaddress]{revtex4-1}
\usepackage{amsmath}
\usepackage{amssymb}
\usepackage{mathtools}
\usepackage{graphicx}
\begin{document}
\title{Topological Lifshitz Transitions, Orbital Currents, and Interactions  \\
in Low-dimensional Fermi Gases in Synthetic Gauge Fields}
\author{Chen-How Huang}
\affiliation{Donostia International Physics Center (DIPC),
Paseo Manuel de Lardizabal, 4. San Sebastian 20018, Spain.}
\author{Masaki Tezuka}
\affiliation{Department of Physics, Kitashirakawa, Kyoto University, Kyoto 606-8502, Japan}
\author{Miguel A. Cazalilla} 
\affiliation{Donostia International Physics Center (DIPC),
Paseo Manuel de Lardizabal, 4. San Sebastian 20018, Spain}
\affiliation{IKERBASQUE, Basque Foundation for Science, E-48011 Bilbao, Spain}
\date{\today}
\begin{abstract}
 Low-dimensional systems of interacting fermions in a synthetic gauge field have been experimentally realized using two-component ultra-cold Fermi gases in optical lattices. Using a two-leg ladder model that is relevant to these experiments, we have studied the signatures of topological Lifshitz transitions and the effects of the inter-species interaction $U$ on the gauge-invariant orbital current  in the regime of large intra-leg hopping $\Omega$. Focusing  on non-insulating regimes, we have carried out numerically exact density-matrix renormalization-group (DMRG) calculations to compute the orbital current at fixed particle number as a function of the interaction strength and the synthetic gauge flux per plaquette. Signatures  of topological Lifshitz transitions where the number Fermi points changes are found to persist even in the presence of very strong repulsive interactions. This numerical observation suggests that  the orbital current can be captured by an appropriately renormalized mean-field band structure, which is also described here.  Quantitative agreement between the mean-field and the DMRG results in the intermediate interaction regime where $U \lesssim \Omega$ is demonstrated. We also have observed that interactions can change the sign of the current susceptibility at zero field and induce  Lifshitz transitions between two metallic phases, which is also captured by the mean-field theory. Correlation effects beyond  mean-field theory in the oscillations of the local inter-leg current are also reported.  We argue that the observed robustness against interactions makes the orbital current a good indicator of the topological Lifshitz  transitions.
\end{abstract}
\maketitle


\section{Introduction}

 Ultracold gases in artificial gauge fields have opened the possibility of quantum simulating condensed matter systems  subject to strong (orbital) magnetic or spin-orbit fields~\cite{RevModPhys.91.015005,Aidelsburger:2014aa_IQHE,PhysRevLett.111.185301_IQHE,PhysRevLett.111.185302_IQHE,PhysRevX.7.021033_laughlinlike,PhysRevB.92.115446_sela_frac,PhysRevB.98.205112_IQHE,PhysRevLett.116.035301}.  In recent experiments~\cite{PhysRevLett.117.220401_sun, Mancini1510_sun, sun_edge,PhysRevLett.118.230402_aeas, PhysRevA.93.061601_aeas, PhysRevLett.116.035301}, it has been possible to simulate fermionic multi-leg ladders subject to a strong (orbital) magnetic field using a Raman laser to couple different internal degrees of freedom (corresponding to different orientations of the nuclear spin of  alkaline-earth-like atoms (AEA)). Thus, the strong inter-leg hopping limit can be achieved by direct control of the Raman lasers. Besides, the Raman lasers provide the atoms a finite momentum kick, which simulates an abelian gauge field. An AEA can be regarded as a spin-$F$ particle where $F$ stands for the nuclear spin. Its $2F+1$ spin states can be coupled by laser-induced complex hopping, leading to a $2F+1$ leg ladders of neutral atoms under a synthetic gauge field as an analogue of 2D systems \cite{PhysRevLett.117.220401_sun, Mancini1510_sun, sun_edge,PhysRevLett.118.230402_aeas, PhysRevA.93.061601_aeas,PhysRevLett.116.035301,Meisner_PhysRevA.102.023315}. Experimentally, several groups have been using synthetic gauge fields in attempts to emulate multi-leg systems in magnetic fields that may display  topological ordered such as  quantum Hall states ~\cite{PhysRevLett.111.185301_IQHE,PhysRevLett.111.185302_IQHE,PhysRevX.7.021033_laughlinlike,PhysRevB.92.115446_sela_frac,PhysRevB.98.205112_IQHE,Haller_2018,Haller_PhysRevResearch.2.023058}, 
which exhibit gapless chiral edge modes~\cite{sun_edge,Stuhl1514_Bose,PhysRevLett.117.220401_sun,Mancini1510_sun,PhysRevLett.112.043001_Rb,Ane1602685_rb}. 

However, to the best of our knowledge, fewer studies have focused on the experimentally realized strong inter-leg hopping limit and experimentally accessible gauge-invariant observables such as the orbital current on which we focus in this article. Unlike  transport~\cite{giamarchi2003,Haller_PhysRevResearch.2.023058}  or (non-quantized) Hall currents~\cite{Buser_PhysRevLett.126.030501} in response to an electric field, the orbital current is a ground state property and  does not depend on the non-equilibrium distribution of  the low-lying degrees of freedom~\cite{PhysRevB.73.195114,PhysRevB.76.195105}. Therefore, in order to understand the effects of interaction it cannot be computed using field-theory methods like e.g. bosonization~\cite{giamarchi2003}. Indeed, the orbital current contains information about the entire band structure and not just the degrees of freedom close to the Fermi energy~\cite{PhysRevB.73.195114,PhysRevB.76.195105}. Furthermore, as we show here, at least in a certain parameter range, the  effects of interactions on this observable can be accounted for by using  mean-field theory. This approach reveals valuable information about the effects of interaction and its interplay with the (synthetic) gauge field.
We have benchmarked the mean-field theory against  density-matrix renormalization group (DMRG) \cite{PhysRevLett.69.2863,PhysRevB.48.10345,Schollwoeck2005,Hallberg2006} calculations of the orbital current.
In the limit of large inter-leg hopping and moderate interaction the mean-field theory can describe the effect of inter-leg interactions.  Qualitatively,  this is because the inter-leg interaction leads to  an effective enhancement of the inter-leg hopping as interactions favor the same type of ground state configurations as the inter-leg hopping. 
 
   As described below, the orbital current displays cusps at certain values of the flux per plaquette or the interaction. In the non-interacting limit, such cusps are related to topological Lifshitz (quantum phase) transitions where the topological properties of the Fermi surface change. In one dimension, the number of disconnected components of the Fermi `surface', i.e. number of Fermi points, changes. Mathematically, this is described by a change in the zeroth order homotopy group, $\pi_0$.  By using DMRG to compute the orbital current, we have found that such cusps are robust in presence of strong interactions.  The mean-field theory described  below is also able to quantitatively describe such cusps for weak to moderate interactions. This suggests that it is possible to interpret the Lifshitz transitions as transitions of a renormalized (mean-field) band structure. 

  Topological Lifshitz transitions have attracted much attention in a variety strongly correlated materials such as heavy fermions~\cite{Vojta_PhysRevB.87.165143}, pnictide superconductors~\cite{pnictides}, and more recently twisted bilayer graphene~\cite{tbg}. In a recent years, Volovik~\cite{volovik1} has  emphasized the importance of Lifshitz
transitions in interacting systems and the non-trivial interplay of topology and interactions, in particular in connection to  flat band formation. In this context, the study reported in this article concerns the effect of Lifshitz transitions in a one-dimensional system  for which interactions are known to have very dramatic effects on the single-particle excitations~\cite{giamarchi2003}. In contrast to this fact,  our findings appear to support the applicability of mean-field theory  to describe the effects of interaction on the orbital current. The mean field theory predicts
 the existence of interaction-driven Lifshitz transitions, which is confirmed by DMRG calculations.

Due to their large tunability, ultra-cold atomic systems appear to be an ideal platform to address some of the deep questions posed by the interplay of interactions and topological Lifshitz transitions. Indeed,  experimental and theoretical work along this direction has been  reported in Ref.~\cite{Wang2012} This study reported the experimental observation of a Lifshitz  transition for a Fermi gas loaded in a quasi-one dimensional optical lattice subject to an artificial spin-orbit coupling. However, unlike the strict one-dimensional regime studied here, the system is not a strongly correlated one, it is not surprising that the results can be described by mean-field theory. In addition, no study of the behavior of the orbital current across the Lifshitz transition  was carried out in Ref.~\cite{Wang2012}. 
 
The following sections are organized as follows: In Sec.~\ref{sec:intro}, we introduce the model and discuss
how the orbital current can be measured in the ultracold atom setup together with the issue
of gauge invariance of observables.  In Sec.~\ref{sec:int}, we discuss the effect of inter-leg interaction using DMRG. The results are compared to perturbation theory and  mean-field theory. We also discuss  the the mean-field theory prediction of an interaction-induced Lifshitz transition, which is confirmed using the results of DMRG for the orbital  and  inter-leg current. 
Finally, we provide a summary and discussion of the results together with an outlook of our work in Sec.~\ref{sec:conclu}.  We have relegated to the Appendices some of most technical details of the calculations as well as the review of some results for non-interacting systems. 

 \begin{figure}[b]
\center
\includegraphics[width=\columnwidth]{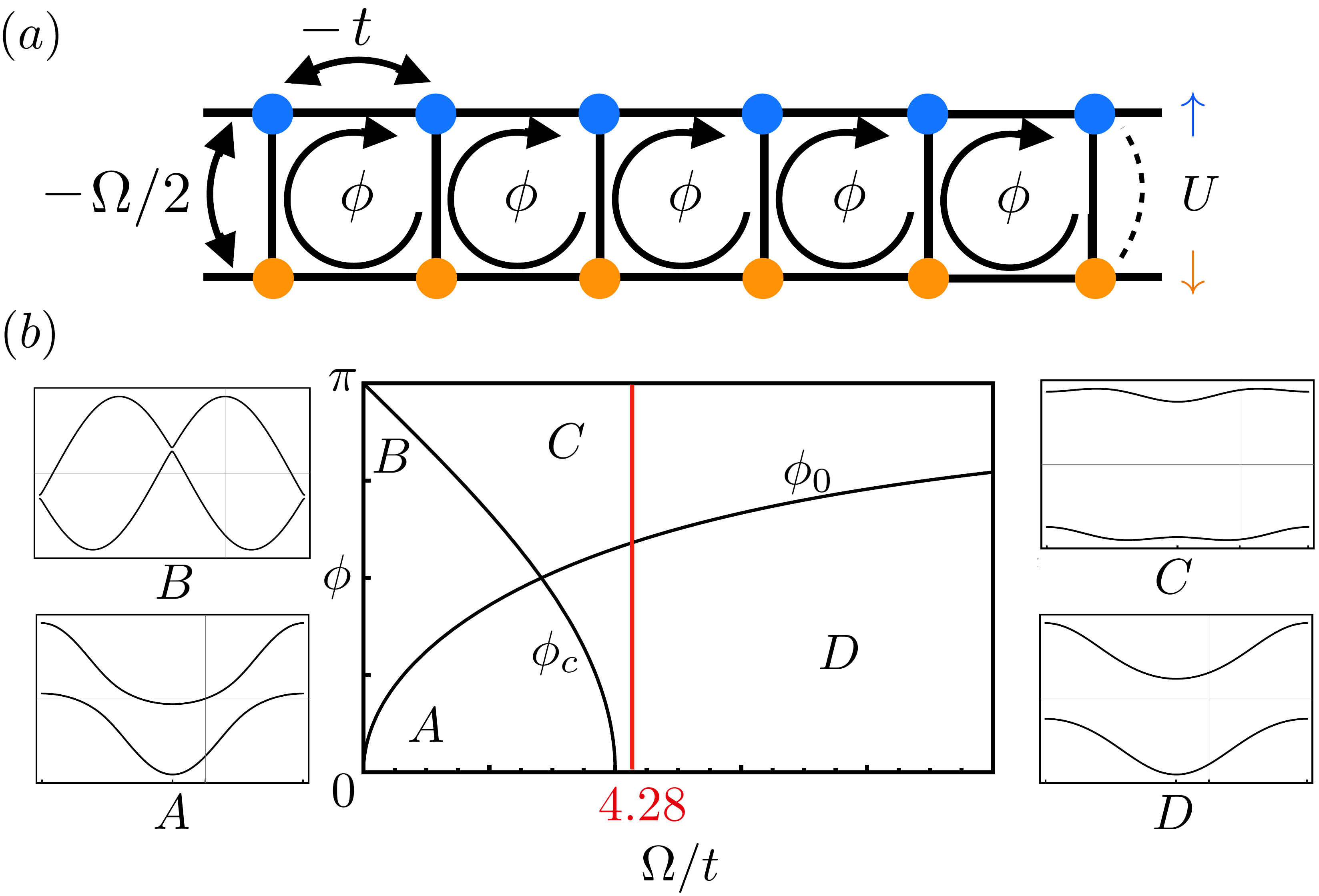}
\caption{(a) Schematic picture of the system considered in this work; $t$ and $\Omega$ parametrize the intra- and inter-leg hopping, and $\phi$ is the applied (synthetic) magnetic flux per plaquette; $U$ parametrizes the strength of the interaction. The leg labels correspond to either $\sigma = \uparrow$ (upper leg) or $\sigma = \downarrow$ (lower leg). (b) The four possible types of band structure  depending on the flux and  the ratio $\Omega/t$ for non-interacting particles (see Appendix~\ref{sec:nonint}). The horizontal (vertical) lines in subfigures A-D correspond to zero  energy  (momentum). For the different types of band structure, we used the convention introduced  in Ref.~\onlinecite{PhysRevB.76.195105}. In the phase diagram,  $\phi_0$ $(\phi_c)$ denotes the appearance of a double-well (band gap). Note that the D phase has only two Fermi points independently of the lattice  filling. However,  the C phase can have more than two, and, in particular, for $\phi=\pi$ it  always has four Fermi points for any lattice filling $n< 1$. The red line corresponds to the  value of $\Omega/t$  realized in the experiment of Ref.~\onlinecite{PhysRevLett.117.220401_sun}. The band structures are centered at wavenumber $q=-(\phi_{\uparrow}+\phi_{\downarrow})/2$. }
 \label{ladder}
\end{figure}

\section{Model and Gauge Invariance}\label{sec:intro}

We shall consider a fermionic two-leg ladder system described by the following Hamiltonian:
\begin{align}
H&= H_{\parallel} + H_{\perp} + H_U, \label{eq:model}\\
H_{\parallel} &= -t\sum_{m,\sigma} \left[\tilde{c}^{\dagger}_{m,\sigma}\tilde{c}_{m+1,\sigma} +\mathrm{H.c.}\right],\\
H_{\perp}&= -\frac{\Omega}{2}\sum_{m} \left[ e^{-i m \phi } \tilde{c}^{\dagger}_{m,\uparrow}\tilde{c}_{m,\downarrow} +\mathrm{H.c.} \right], \label{eq:hperp}\\
H_{U} &=U\sum_{m}\tilde{n}_{m,\uparrow}\tilde{n}_{m,\downarrow}.\label{eq:int}
\end{align}
In the above expression $t$ ($\Omega$) is the intra-leg  (inter-leg) hopping,
and $\tilde n_{m\sigma} = \tilde c^{\dag}_{m\sigma} \tilde c_{m\sigma}$. Anticipating that in ultracold atomic systems  the legs of the ladder may correspond to different internal degrees of freedom or to   the excited and ground states of the same atom species~\cite{PhysRevLett.117.220401_sun}, we use a pseudo-spin index $\sigma = \uparrow, \downarrow$ that refers to the chain index. In this notation, the orbital current that we discuss here  can also be regarded as a pseudo-spin current (cf. Eq.~\ref{eq:orbcurrent}).  The interaction of the Hubbard type acts on  atoms with different pseudo-spin or leg index $\sigma$.  Concerning this point, we recall that, at ultracold temperatures,  interactions are well described by a short-range pseudo-potential which leads to Eq.~\eqref{eq:int} when projected on the Wannier orbitals of the lowest Bloch bands.
In the above expression, $\phi$ corresponds to the flux of the Aharonov-Bohm phase of pseudo-gauge field accumulated by an atom when going around a plaquette (see Fig.~\ref{ladder}).  Finally, note that in Eq.~\eqref{eq:model} and in all the equations below,  we work in units where the lattice constant $a=1$  and the reduced Planck's constant $\hbar=1$.

  Let us perform a (unitary) gauge transformation such that 
\begin{equation}
c_{m,\sigma}= \tilde{c}_{m,\sigma} e^{i m \phi_{\sigma} }. \label{eq:gaugetr}
\end{equation}
By introducing  $\phi_{\uparrow}$ and $\phi_{\downarrow}$ obeying $\phi_{\downarrow}-\phi_{\uparrow}=\phi$,  the site dependence in the phase of the inter-leg hopping term, $H_{\perp}$, can be removed at the cost of introducing 
 (position-independent) phases in the intra-leg hopping terms of the Hamiltonian, i.e.
 \begin{align}
 H_{\parallel} &= -t \sum_{m,\sigma} \left[ e^{i\phi_{\sigma}} c^{\dag}_{m,\sigma} c_{m+1,\sigma} + \mathrm{H.c.} \right] \\
 H_{\perp} &= -\frac{\Omega}{2} \sum_{m} \left[c^{\dag}_{m,\uparrow} c_{m,\downarrow} + \mathrm{H.c.} \right].
 \end{align}
 Note that the gauge transformation~\eqref{eq:gaugetr} 
 does not affect the site occupation operators, i.e. $n_{m,\sigma} = c^{\dag}_{m,\sigma} c_{m,\sigma} =  \tilde{c}_{m,\sigma} \tilde{c}_{m,\sigma}  = \tilde{n}_{m,\sigma}$, and therefore the form of the interaction  is not altered.
 Let us recall that,  in condensed matter systems, all observables
 must be gauge invariant and cannot depend on a 
 particular choice of the phases $\phi_{\sigma}$.  On the other hand,
 in ultracold atomic systems, non-gauge invariant
 observables are experimentally accessible. An important 
 example of such non-gauge invariant observable is the
 chirality used in the experiment of Refs.~\onlinecite{Mancini1510_sun,PhysRevLett.117.220401_sun}
to detect the presence of  chiral currents induced by the synthetic gauge field.
In the setup of Refs.~\onlinecite{Mancini1510_sun,PhysRevLett.117.220401_sun} a particular gauge choice
where $\phi_{\uparrow} = \phi$ and $\phi_{\downarrow} = 0$ 
is realized. In order to detect the presence of chiral 
currents, the chirality is measured. The latter is mathematically defined as:
\begin{equation}
C= \int _{-\pi}^{+\pi} \frac{dq}{2\pi}\, n_{\uparrow}(q) \text{sign}(q),
\label{eq:chirality}
\end{equation}
where  $n_{\uparrow}(q) =   \langle c^{\dag}_{q\uparrow} c_{q\uparrow} \rangle$ (with $c_{q,\sigma}  = L^{-1/2} \sum_{m} e^{iq m} c_m$) is the lattice momentum distribution of the $\uparrow$ fermions. Note that the momentum distribution $n_{\uparrow}(q)$  and therefore the chirality $C$ in Eq.~\eqref{eq:chirality} are not invariant under  gauge transformations such like Eq.~\eqref{eq:gaugetr}. This means that they have no counter-part in  condensed matter systems which could be emulated with this atomic system.

 However, a gauge invariant quantity can be defined and measured if we replace the  $\text{sign}(q)$ function  in Eq.~\eqref{eq:chirality} by a (shifted) sinus function:
\begin{equation}
\tilde{C}(\phi)= \int _{-\pi}^{+\pi} \frac{dq}{2\pi}\,  n_{\uparrow}(q) \sin(q+\phi).
\end{equation}
In the gauge choice used by the experiment this quantity 
is  related to the derivative of the energy with respect to the
flux $\phi_{\uparrow} = \phi$ (cf. Eq.~\ref{eq:orbcurrent})
\begin{equation}
 2t \tilde{C}(\phi) = \frac{1}{L} \left\langle\frac{\partial H}{\partial\phi_{\uparrow}}\right\rangle = \frac{1}{L} \left\langle\frac{\partial H}{\partial\phi}\right\rangle.
\end{equation}
The latter derivative is gauge invariant and therefore does not depend on $\phi_{\uparrow}$ and $\phi_{\downarrow}$ separately but only on their difference $\phi = \phi_{\uparrow} - \phi_{\downarrow}$. Indeed,  for a general gauge  choice, it is proportional to the orbital current per unit length,
\begin{equation}
\frac{J(\phi)}{L}=  -\frac{1}{L} \left\langle
\frac{\partial H}{\partial\phi} 
\right\rangle = -2t \tilde{C}(\phi),
\end{equation}
which is the observable on which we shall focus our attention in this work.  Thus, we emphasize that,  since chirality is can be measured in the experiments of Refs.~\cite{Mancini1510_sun,PhysRevLett.117.220401_sun}, the orbital current is also measurable. Indeed,  both rely on the measurement of  the momentum distribution of the atoms in the $\sigma=\uparrow$ internal state in the  gauge choice implemented  in Refs.~\cite{Mancini1510_sun,PhysRevLett.117.220401_sun}.
 \section{Orbital Currents}\label{sec:eff}
%
In ultracold atom systems, unlike many condensed matter systems,
as the flux per plaquette $\phi$ is varied, the particle
number (and not  the chemical potential) remains constant. 
Thus, in this section,  we describe the 
calculation  of the orbital current and the orbital susceptibility at fixed particle number.  Although in 
our numerical and analytical calculations we have restricted ourselves to zero temperature, it is useful to discuss the
finite temperature case. In what follows, 
we set Boltzmann's constant $k_\mathrm{B}=1$ and denote the inverse absolute temperature by $\beta$ ($= T^{-1}$). 

 Let us first consider  the partition function in the grand canonical ensemble:
\begin{align}
Z(\beta,\mu,\phi) &=\text{Tr}\left[e^{-\beta \left( \hat{H}(\phi) -  \mu \hat{N}\right)}\right],
\end{align}
where $\mu$ is the chemical potential. In condensed matter systems at fixed $\mu$, 
the orbital current is defined as a derivative of the grand canonical potential, i.e.
\begin{align}
\Omega(\beta,\mu,\phi) &= -\frac{1}{\beta} \log Z(\beta,\mu,\phi),\\
J(\mu,\phi) &=  \left(\frac{\partial\Omega(\beta,\mu,\phi)}{\partial \phi}\right)_{\beta,\mu} \\ 
&= \frac{\mathrm{Tr} \left[e^{-\beta\left( \hat{H}(\phi) - \mu \hat{N}\right)}  \hat{J}(\phi)\right]}{Z(\beta,\mu,\phi)} \\
&= \langle \hat{J}(\phi) \rangle_{\mu},
\end{align}
where 
\begin{align}
\hat{J}(\phi) &= - \frac{\partial \hat{H}(\phi)}{\partial \phi}\label{eq:Jop}\\
&=  it \sum_{m,\sigma}  \left(\frac{d\phi_{\sigma}}{d\phi} \right) \left[ e^{i  \phi_{\sigma}} c^{\dag}_{m,\sigma} c_{m+1,\sigma} - \mathrm{H.c.} \right] \\
&= -2t\sum_{q\in \mathrm{1BZ},\sigma}  \left(\frac{d\phi_{\sigma}}{d\phi} \right)  
\sin \left(q + \phi_{\sigma}\right) \,  c^{\dag}_{q\sigma} c_{q\sigma}    \label{eq:orbcurrent}
\end{align}
is the  orbital current operator and  $\mathrm{1BZ}$ stands for the 1st Brillouin zone, which corresponds to the segment $(-\pi,\pi ]$. 

 However, we are interested in the current $J(N,\phi)$ at fixed particle number $N$. In order to obtain the latter, we can start from the constraint equation for the particle number $N$:
\begin{equation}
N = \langle \hat{N}\rangle_{\mu} = - \left(\frac{\partial\Omega(\beta,\mu,\phi)}{\partial \mu} \right)_{\beta,\phi} = N(\beta,\mu,\phi).
\end{equation}
By solving this equation for $\mu = \mu(\beta,N,\phi)$ and introducing the solution
into $\tilde{J}(\beta,\mu,\phi)$ we  obtain  the orbital current  $J(\beta, N, \phi)$ as a function of  $N$. 

 Alternatively, we can work  with the thermodynamic potential that is the Legendre transform of the Grand canonical potential $\Omega(\beta,\mu,\phi)$, i.e.
\begin{equation}
G(\beta,N,\phi) = \Omega(\beta,\mu(\beta,N,\phi), \phi) + \mu(\beta,N,\phi) N
\end{equation}
Hence,  the orbital current for a fixed particle number can be obtained from the free energy,
\begin{equation}
 G(N,\phi) = -\frac{1}{\beta}\log Z(\phi) + \mu(\phi) N 
\end{equation}
as a derivative:
\begin{align}
J(N,\phi)&=-\left[\frac{\partial}{\partial \phi}G(\beta,N,\phi)\right]_{\beta,N} \\
&=\frac{\text{Tr}\left[ e^{-\beta \left(\hat{H}(\phi) - \mu(\beta,N,\phi)\hat{N} \right)}\hat{J}(\phi)\right]}{Z\left(\beta,\mu(\beta,N,\phi),\phi\right)} \\
 &= \langle \hat{J}(\phi) \rangle_{N}.  \label{eq:JJ}
\end{align}
Another quantity of interest is the orbital susceptibility, which is defined from the derivative
of the orbital current density with respect to $\phi$:
\begin{align}
\chi(\beta,N, \phi) &=  \left( \frac{\partial}{\partial \phi} \frac{J(\beta,N,\phi)}{L} \right)_{\beta,N} \\
&= - \frac{1}{L}\left[\frac{\partial^2}{\partial \phi^2}G(\beta,N,\phi)\right]_{\beta,N}.
\end{align}
Note that  the free energy $G(\beta, N, \phi)$ reduces to the ground state energy calculated at fixed particle number $N$  at the zero temperature.  Likewise, the expectation of Eq.~\eqref{eq:Jop} is taken over the ground state of the system containing 
$N$ particles. In the DMRG calculations of $J(N,\phi)$   that are described in the following section, the constant particle number constraint is imposed by projecting the ground state on the subspace of the Hilbert space with total particle number equal to $N$ (the latter is determined by the system length $L$ and the lattice filling $n = N/L$).

\begin{figure}[b]
\includegraphics[width=\columnwidth]{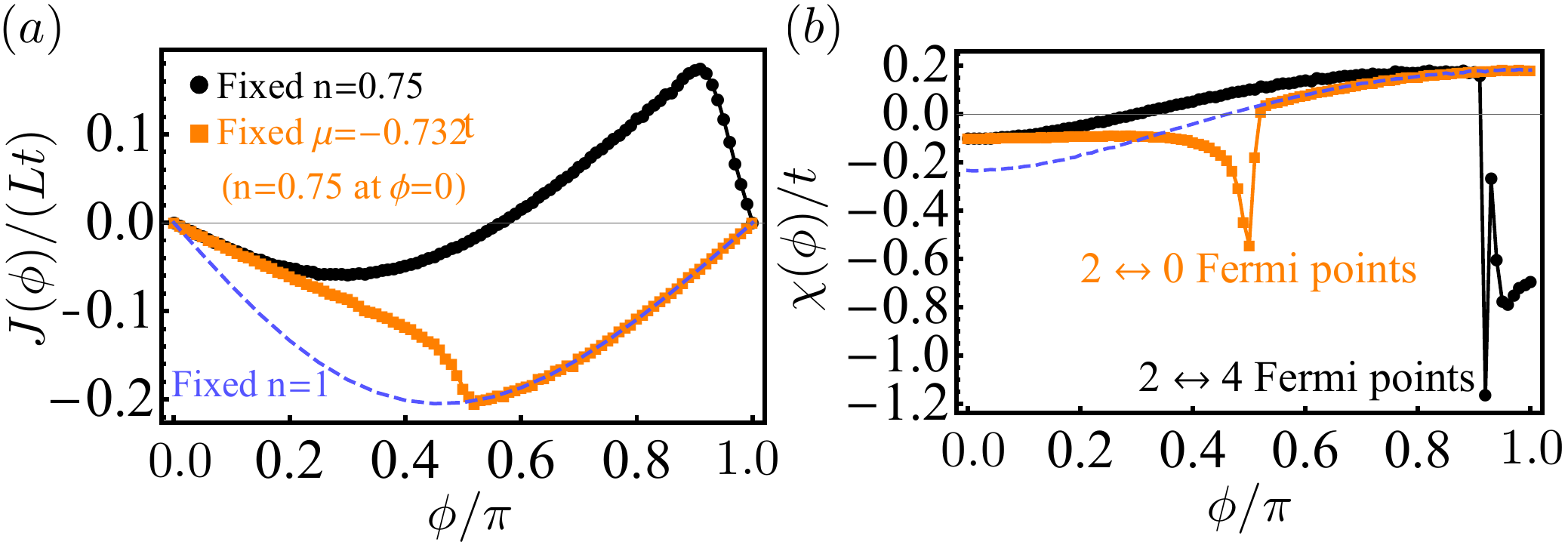}
\caption{ Panel (a): Comparison of the orbital current obtained at fixed particle number 
(corresponding to a lattice filling of $n=0.75$) and fixed chemical potential ($\mu/t=-0.732$, which is the chemical potential resulting in a lattice filling of $n=0.75$ at $\phi=0$). Note that in both cases the current exhibits a cusp behavior with a discontinuous derivative due to the changes in numbers of Fermi points in the occupation of the bands. The current at fixed lattice filling $n=1$, which corresponds to a completely filled lower band, is shown
to illustrate that, at fixed chemical potential, the system undergoes a transition for  $\phi/\pi \simeq 0.55\pi$  from a metal to a band insulator where the number of Fermi points drops from two to zero.  Panel (b) shows a comparison of the  orbital susceptibility at fixed $n$ and fixed $\mu$ as corresponding to the parameters used in panel (a). Note 
that the orbital susceptibility exhibits a divergence followed by an abrupt sign change (as corresponds to the cusps observed in the current) at the transition where the number of Fermi points changes. In all plots we have set the ratio $\Omega/t = 4.28$, which is the value used in the experiment of Ref. ~\cite{PhysRevLett.117.220401_sun}.}\label{fig:jmu}
\end{figure}
\begin{figure}[b]
\includegraphics[width=\columnwidth]{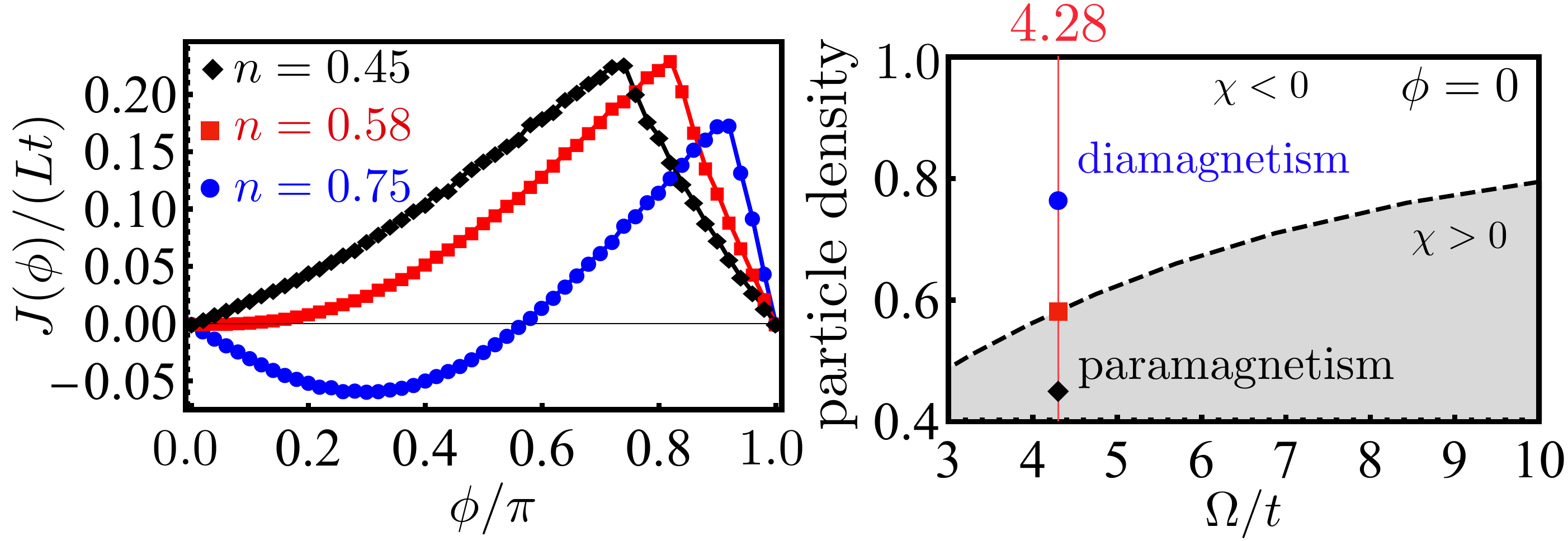}
\caption{ (a) Ground state value of the orbital current density for the non-interacting  system as a function of the applied pseudomagnetic flux per plaquette, $\phi$ for different lattice and $\Omega/t=4.28$. At small $\phi$, the current shows diamagnetic/paramagnetic dependence on the flux. At $\phi$ near $\pi$, the current shows cusp discontinuity which is related to the change in the number of Fermi points.
 (b)  "Phase diagram" of the zero field ($\phi=0$) orbital susceptibility. The behavior of the latter is determined by the particle density and the ratio of hopping amplitudes $\Omega/t$. In the calculation shown in this diagram we assumed that $\Omega/t\gg1$ is small enough for the contribution from the upper band to be neglected. The points indicated on the figure correspond to the fillings used in the calculations shown on the left panel.}\label{fig:j0}
\end{figure}

To illustrate the differences between the orbital current and susceptibility computed at fixed particle number and  chemical potential, we have  plotted them in Fig.~\ref{fig:jmu}. The chemical potential at $\phi = 0$ has been chosen to yield a  lattice filling of $n = 0.75$, which is the same value used for the calculations at fixed particle number. Thus, although the curves for the orbital current and the susceptibility are very close at small $\phi$, they show large deviations
for $\phi \gtrsim \frac{\pi}{4}$. This very different behavior can be understood from the  changes in the band occupation in the cases of fixed $n$ and fixed $\mu$: Whereas at fixed $n  =0.75$ the system transitions from a metal with two Fermi points to a metal with four Fermi points, for constant $\mu$ the system transitions from a metal to a band insulator with the chemical potential moving into the band gap as $\phi$ changes from $0$ to $\pi$. At the transition points,  the orbital current shows a cusp and the susceptibility becomes singular. However,  beyond the cusp,  the curves at fixed $\mu$ for both the orbital current and the susceptibility fall on top of the curves  with constant lattice filling $n = 1$ (dashed curves). For unit lattice filling,  the system is a band insulator at any $\phi$ for large $\Omega$. Thus, to sum up, whilst the system at fixed lattice filling exhibits a Lifshitz transition between two different metallic phases, for fixed  chemical potential  it undergoes a transition from a metal with two Fermi points to a band insulator. The singular behavior of the susceptibility is  different for the two kinds of constraints. The different types of singularities are  analytically calculated in  Appendix~\ref{app:constraint}. We find that, for a non-interacting system with fixed particle number,  the orbital susceptibility has   a step-like discontinuity. On the other hand, $(\phi-\phi_c)^{-1/2}$ singularities are found for  fixed chemical potential.

   To further explore  the dependence on the lattice filling in the non-interacting case, in Fig.~\ref{fig:j0} we have plotted the  current at different values of $n$. Note that the orbital current vanishes both at $\phi = 0$ and $\phi = \pi$. In the latter case, the  hopping amplitudes are all real  but the unit cell size doubles (To see this,  notice that, e.g. in the gauge choice of Eq.~\eqref{eq:model}, the phase of the inter-leg hopping in $H_{\perp}$, Eq.~\eqref{eq:hperp} becomes $(-1)^j$ for $\phi = \pi$). In addition, in all cases shown  in Fig.~\ref{fig:j0},   the  orbital current reaches a maximum and exhibits a cusp as $\phi$ changes from $0$ to $\pi$. As for the case where $n =0.75$ studied above,  this happens when the system undergoes a topological Lifshitz transition where the number of Fermi points changes. The transition is also reflected in the orbital susceptibility as a divergence (see Appendix~\ref{app:constraint}).  
Below, we shall see that the cusp in the orbital current persists in the presence of interactions. This observation strongly suggest that at least some of the interaction effects can be captured by a properly renormalized (mean-field) band structure as we show in Sec.~\ref{sec:mft}.

\section{Interaction Effects}\label{sec:int}

\subsection{Limit of large inter-leg hopping}\label{sec:largeomega}

The two-leg ladder model introduced in Sec.~\ref{sec:intro} contains three different
energy scales $t, \Omega$ and $U$. In order to appreciate the importance of the interaction in
 the limit where $\Omega$ is the largest  of the three energy scales, let us fix first 
 the Gauge to the symmetric gauge where $\phi_{\uparrow} = -\phi_{\downarrow}  = \phi/2$. 
 For large $\Omega > 0$,  it is  convenient to diagonalize the inter-leg hopping
 term by means of a unitary transformation:
\begin{equation}
\left(
\begin{array}{c}
c_{m,\rightarrow}\\
c_{m,\leftarrow}
\end{array} \right)
= \frac{1}{\sqrt{2}} \left(
\begin{array}{cc} 
1 &1 \\
1 & -1
\end{array} \right) 
\left(
\begin{array}{c}
c_{m,\uparrow}\\
c_{m,\downarrow}
\end{array} \right).
\end{equation}
Thus, 
\begin{equation}
H_{\perp} = -\frac{\Omega}{2} \sum_{m} \left( c^{\dag}_{m,\rightarrow} c_{m,\rightarrow} - c^{\dag}_{m,\leftarrow} c_{m,\leftarrow} \right)
\end{equation}
and the lowest energy state at a given site $m$ is $|\rightarrow\rangle_m = c^{\dag}_{m,\rightarrow}|0\rangle$, 
which is separated by an energy of $\Omega$ from the $|\leftarrow\rangle_m =  c^{\dag}_{m,\leftarrow}|0\rangle$ state. 

 Let us next consider the interaction term,  $H_U$. When performing the above spin rotation, it is  useful to write 
it in a rotational  invariant fashion, in terms of the site occupation $n_{m} = n_{m,\uparrow} + n_{m,\downarrow} = 
n_{m,\rightarrow}+n_{m,\leftarrow}$. Hence, it follows that $H_U$ remains unchanged:
\begin{align}
H_U &= U\sum_{m} n_{m,\uparrow} n_{m,\downarrow} \\
&= \frac{U}{2} \sum_{m} \left(n_{m} -1 \right)^2 + \frac{U}{2} N - U  \\
 &= U \sum_{m} n_{m,\rightarrow} n_{m,\leftarrow}
\end{align}
For  $t=0$ and lattice filling $n \leq 1$, the above expressions imply that  repulsive
interactions play  no role in the ground state as all  sites can be
occupied either with a single fermion in the $|\rightarrow\rangle_m$ level or an empty site (i.e. $|0\rangle_m$).
Thus, as long as $n \leq 1$ and $t=0$, the many-particle states in this subspace 
are all degenerate in energy. This large degeneracy is lifted by the hopping term,  $H_{\parallel}$, which, in the new basis, takes the  form:
\begin{align}
H_{\parallel} &= -  \sum_{m}  \left( \begin{array}{cc}
c^{\dag}_{m+1,\rightarrow} c^{\dag}_{m+1,\leftarrow} 
\end{array} \right)  \mathcal{T}(\phi)
\left(
\begin{array}{c}
c_{m,\rightarrow}\\
c_{m,\leftarrow}
\end{array} \right) \notag\\
&\qquad + \mathrm{h.c.} \\
\mathcal{T}(\phi)  &= t
 \left( 
\begin{array}{cc}
\cos \phi/2 & i \sin \phi/2 \\
i\sin \phi/2 & \cos \phi/2 
\end{array} 
\right)
\end{align}
Note that the price to pay  for working with $\{ c_{m,\rightarrow}, c_{m,\leftarrow} \}$ is a non-diagonal 
$H_{\parallel}$. For fermions in the low-lying $\rightarrow$ level and as long as $t\ll \Omega$, 
the hopping amplitude becomes $t\cos\phi/2$ and we are left with 
a band with a modified dispersion:
\begin{equation}
\tilde{\epsilon}_{q} = -2 t \cos (\phi/2) \cos q. \label{eq:appdis}
\end{equation}
Thus, we can regard the system at filling $n < 1$ as 
a ferromagnetic (or spin-polarized) metal which is a weakly interacting Fermi  gas 
as long as $U\lesssim \Omega$. The weak effective interactions are introduced by virtual hopping processes
mediated by the off-diagonal hopping
terms of $H_{\parallel}$, which admix the $\rightarrow$ and $\leftarrow$ levels 
thus allowing the fermions to interact.  Note that the system does not have a charge gap but it has a spin gap $\sim \Omega > U$,
which is the energy cost to flip one spin from $\rightarrow$ to $\leftarrow$. Thus, with the possible exception 
of half-filling, we can regard the system as a spin-polarized weakly correlated
Tomonaga--Luttinger liquid~\cite{giamarchi2003} 
(see also discussion in Sect.~\ref{sec:strongU} below).

Nevertheless,   the simple picture provided above runs into trouble
because the (lowest) band flattens for $\phi\to \pi$  since its width decreases as   $w = 4 t \cos \phi/2$.
At the same time, the magnitude of the 
off-diagonal spin-flip hopping   terms in $H_{\parallel}$ increases. This  results
in an increase of the admixture between the $\rightarrow$ and $\leftarrow$ levels.
Thus,   
for $\phi \gtrsim \pi/2$, interaction effects are expected to become 
more important. This 
expectation is confirmed by the DMRG results  for the orbital current 
discussed below (see e.g. 
Fig.~\ref{fig:DMRGn025} and the discussion in the next subsection).

 Nevertheless, we also find instances at small $\phi$ for which the
behavior of the orbital current can be substantially altered by interactions 
(cf. Fig.~\ref{fig:DMRGn075}). This can be understood as follows: Even 
in the absence of interactions,  the orbital susceptibility at small $\phi$  is not entirely 
determined by the dispersion in \eqref{eq:appdis} but also depends on the
admixture of the  $\rightarrow$ and $\leftarrow$ levels. This admixture is
responsible for the sign of the orbital susceptibility  $\phi = 0$ (see Fig. \ref{fig:j0} as well as 
Eq.~\eqref{eq:chi0} in  Appendix~\ref{app:diapara}). As noted above,  the admixture also 
allows fermions to interact. Therefore,  as $U$ is increases,
the orbital susceptibility at small $\phi$ gets renormalized in a rather dramatic way  
by changing sign: the system  undergoes  a transition from 
diamagnetic to paramagnetic behavior (cf. Fig.~\ref{fig:DMRGn075}).
Note that such change of behavior is suppressed by a sufficiently large $\Omega$ as also
shown on the right panel of  Fig.~\ref{fig:DMRGn075}.  

\subsection{DMRG Results}
\begin{figure}[ht]
\includegraphics[width=\columnwidth]{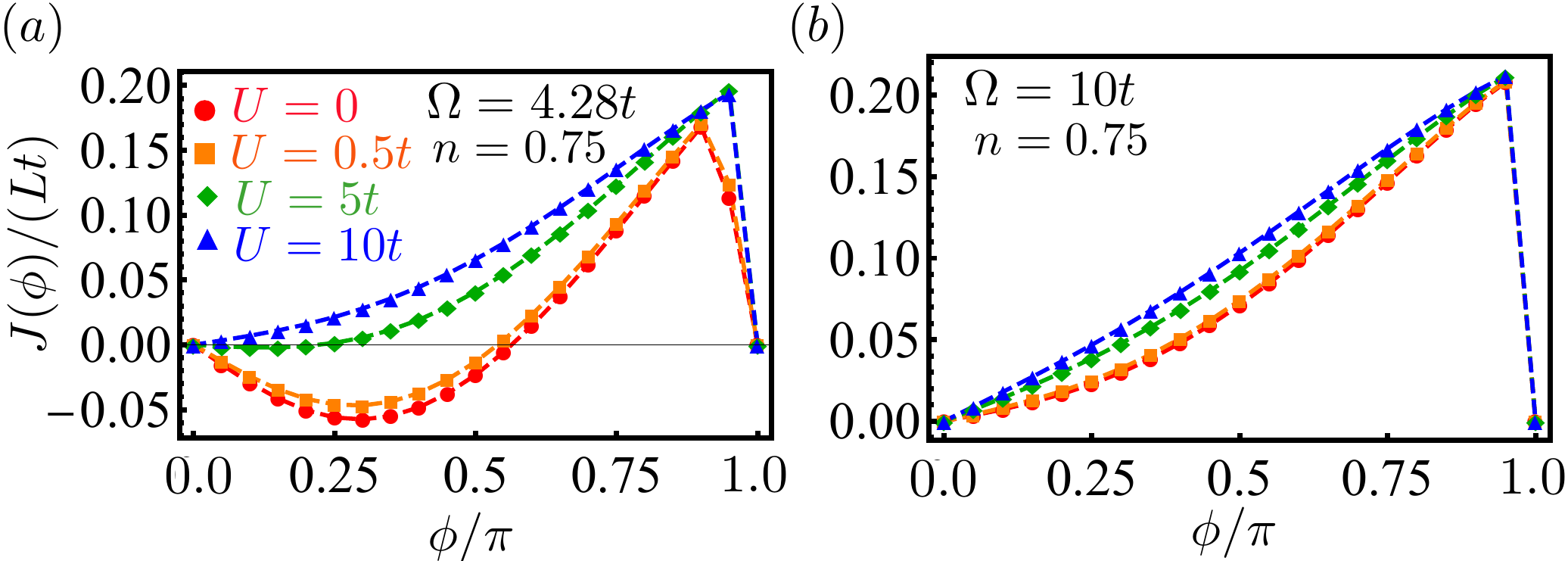}
\caption{Effect of inter-chain (i.e. inter-species) interaction on the orbital current as calculated using DMRG.  The lattice filling is $n=0.75$ and two values of the inter-chain hopping are shown: $\Omega = 4.28 t$, which is the value in the experiment reported in
Ref.~\cite{PhysRevLett.117.220401_sun} and $\Omega = 10 t$. DMRG results are shown as  dots. The dashed lines are a guide to the eye.  Panel (a) shows that interactions  can drive a transition from diamagnetic to paramagnetic behavior.  We define diamagnetic (paramagnetic) behavior by the negative (positive) orbital susceptibility at $\phi = 0$, i.e. the slope of the orbital current at small $\phi$.  Note that, as shown on panel (b),  no paramagnetic-diamagnetic transition occurs for the larger value of $\Omega=10t$. The overall effect of interaction becomes weaker for larger $\Omega$.}\label{fig:DMRGn075}
\end{figure}
\begin{figure}[hb]
\includegraphics[width=\columnwidth]{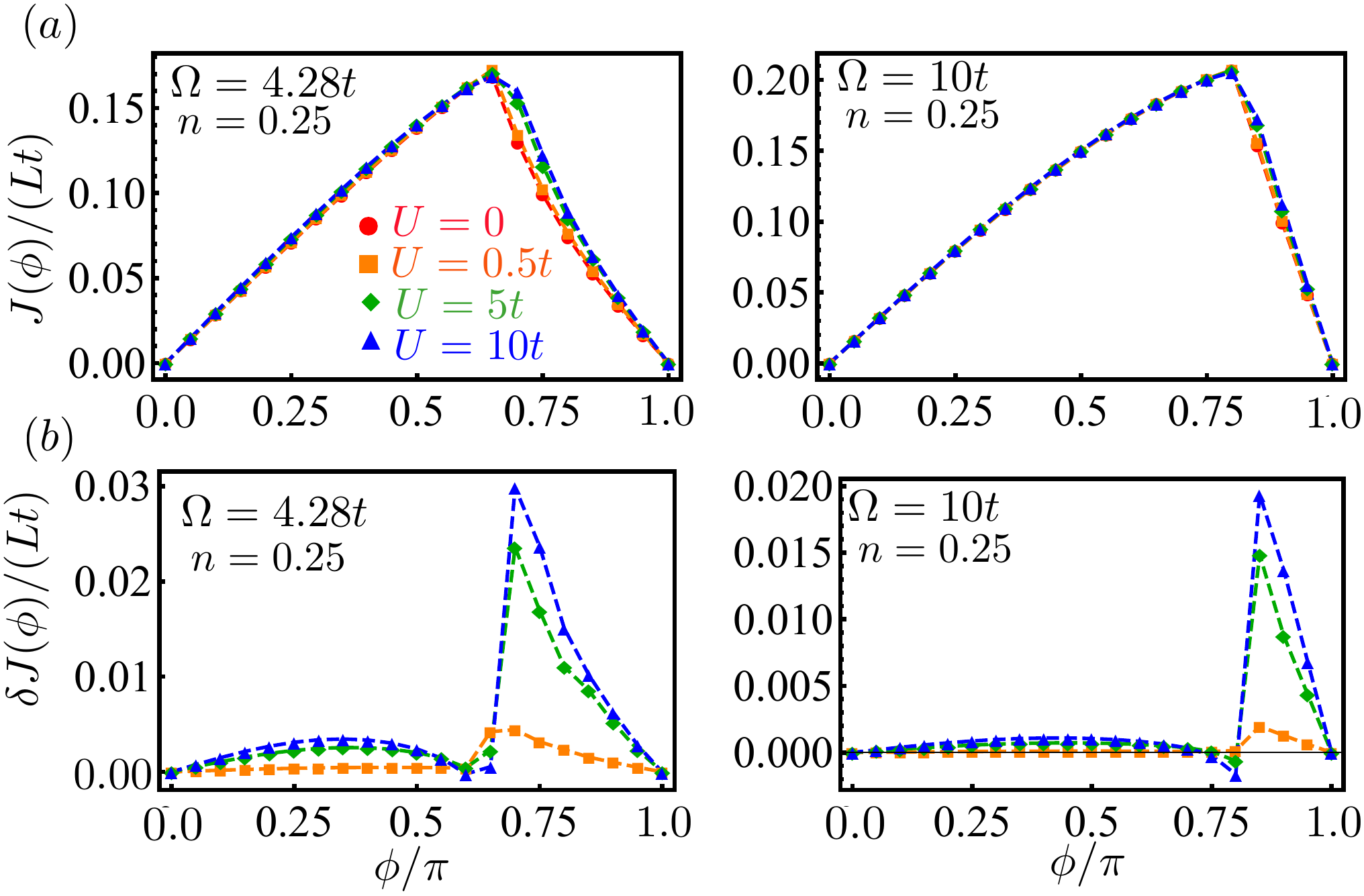}
\caption{ DMRG results for the orbital current (upper panels, a) and the deviation from the non-interacting orbital current (lower panels, b) for several values of the inter-leg interaction $U$ and the inter-leg hopping  $\Omega$ with lattice filling $n=0.25$. (a)  At this lower lattice filling, the overall effect of inter-leg interaction on the orbital current is weaker and, in agreement with the results shown Fig.~\ref{fig:DMRGn075},  it is  further suppressed at larger $\Omega/t$.  To  show the  effects of the interaction more clearly, we have also plotted the deviation of the orbital current from the non-interacting limit. Note the interaction effects are more important for $\phi$ larger than the flux value for which the orbital current exhibits a cusp.}\label{fig:DMRGn025}
\end{figure}
\begin{figure}[b]
\includegraphics[width=0.9\columnwidth]{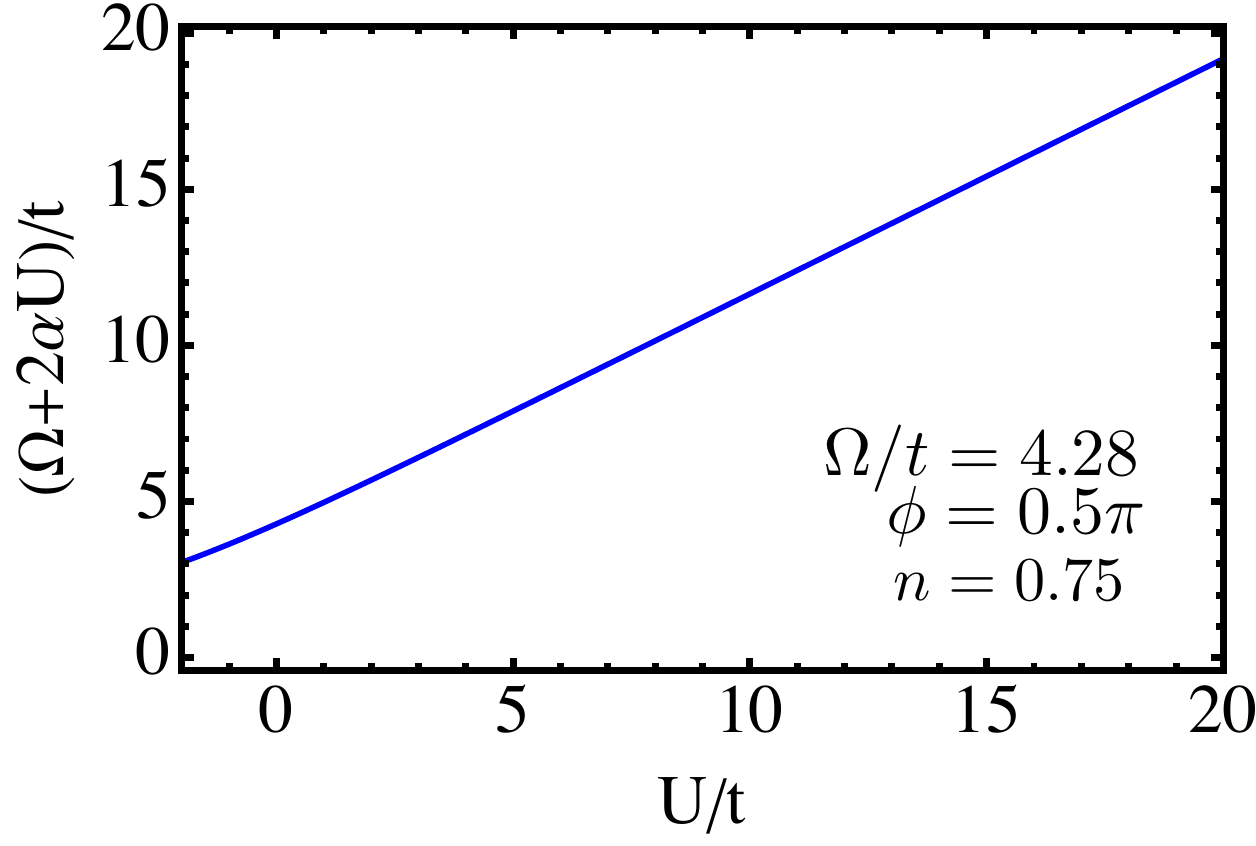}
\caption{(Average) mean-field renormalized inter-leg hopping $\Omega$ as a function of the inter-leg strength $U/t$. In a system with open boundary conditions, the order parameter $\alpha_m(\phi) = \langle c^{\dag}_{m,\uparrow}c_{m,\downarrow}\rangle$ is not uniform. Here we have plotted the average $\alpha = \sum_{m} \alpha_m(\phi)/L$.}\label{fig:renomega}\label{fig:omegamft}
\end{figure}
 We have carried out DMRG calculations using the ALPS Library \cite{ALPS2011, ALPS2007} for systems up to $L=128$ sites, using up to $128$ block states. Details about the numerical convergence of the  results are provided  in  Appendix~\ref{app:DMRG}.  Here suffice to mention that, in the non-interacting limit for which analytical results in the thermodynamic limit can be obtained, the difference between the orbital current in thermodynamic limit and the  DMRG results obtained for systems of $L=128$ sites is within $1$\%.

 Figs.~\ref{fig:DMRGn075} and \ref{fig:DMRGn025} show our DMRG results for the orbital current for two values of the inter-leg hopping $\Omega/t = 4.28$ and $\Omega/t = 10$ ($\Omega/t = 4.28$ corresponds to the experimental value~\cite{PhysRevLett.117.220401_sun}), two values of the lattice filling, $n=0.75$ and $n = 0.25$ for $U$ ranging from zero to $10t$.  Note the presence of a cusp in all curves of the orbital current as $\phi \to \pi$. The latter is already present in the non-interacting case (i.e. $U = 0$) and corresponds, as explained above, to a topological transition where the number of Fermi points  in the lower band  changes from two to four Fermi points. Generally,  the position of the cusp is pushed towards slightly larger values of $\phi$ as the strength of  $U$ increases. This effect is more noticeable for the smaller  $\Omega$ and  the larger lattice filling $n$,  for which interactions have a more pronounced effect (see below).
 
 Another noticeable effect is that repulsive interactions tend to  enhance the orbital current.  This is precisely the opposite of what is expected for  attractive interactions, for which fermions on different legs (i.e. different pseudo-spins) tend to form pairs for sufficiently large negative $U$, which results in a suppression of the orbital (or pseudo-spin) current. Furthermore,  as mentioned above,  for the smaller value of $\Omega/t$ and for small values of  $\phi$, this  enhancement can actually  reverse the slope of the orbital current at $\phi = 0$,  thus inducing  a transition from  ``paramagnetic'' behavior (characterized by an orbital susceptibility $\chi < 0$ at zero field) to  ``diamagnetic"  behavior (for which $\chi > 0$ at $\phi= 0$), see Figs.~\ref{fig:j0} and \ref{fig:DMRGn075} and Appendix~\ref{app:diapara}.   Fig.~\ref{fig:DMRGn025} shows the orbital current and its deviation from the non-interacting case using DMRG computations at low filling $n=0.25$. In the limit of low density and large $\Omega$, the effect of interaction is weak ($<10\%$ for $U$ up to $10t$). Note as well that the orbital current appears to be more sensitive to interaction effects  beyond the cusp. The observed trends in the DMRG results can be qualitatively understood from the following  reasoning: Since the  interaction energy per unit length is of order $\sim Un^2/4$ whereas in the large $\Omega$ limit the kinetic  term is $\sim \Omega n+O(t)$, the relative importance of the interaction can be estimated from their ratio $\sim Un/\Omega$, which decreases with lattice filling and larger $\Omega$.


%
\begin{figure}[b]
\includegraphics[width=\columnwidth]{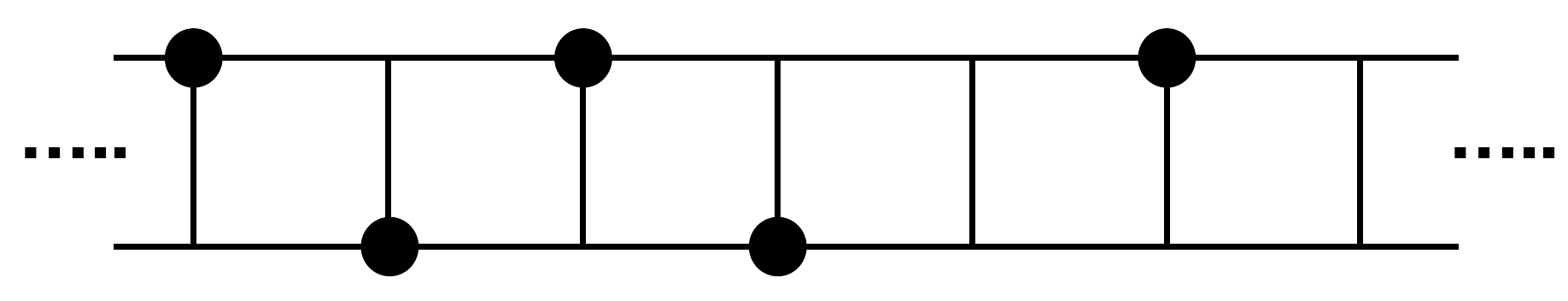}
\caption{One of the possible ground-state configurations in the large $U > 0$ and large inter-leg hopping, where all the occupied sites have no neighboring sites in the other chain occupied.}\label{fig:mf}
\end{figure}

\subsection{Mean-field theory}\label{sec:mft}
\begin{figure}[t]
\includegraphics[width=\columnwidth]{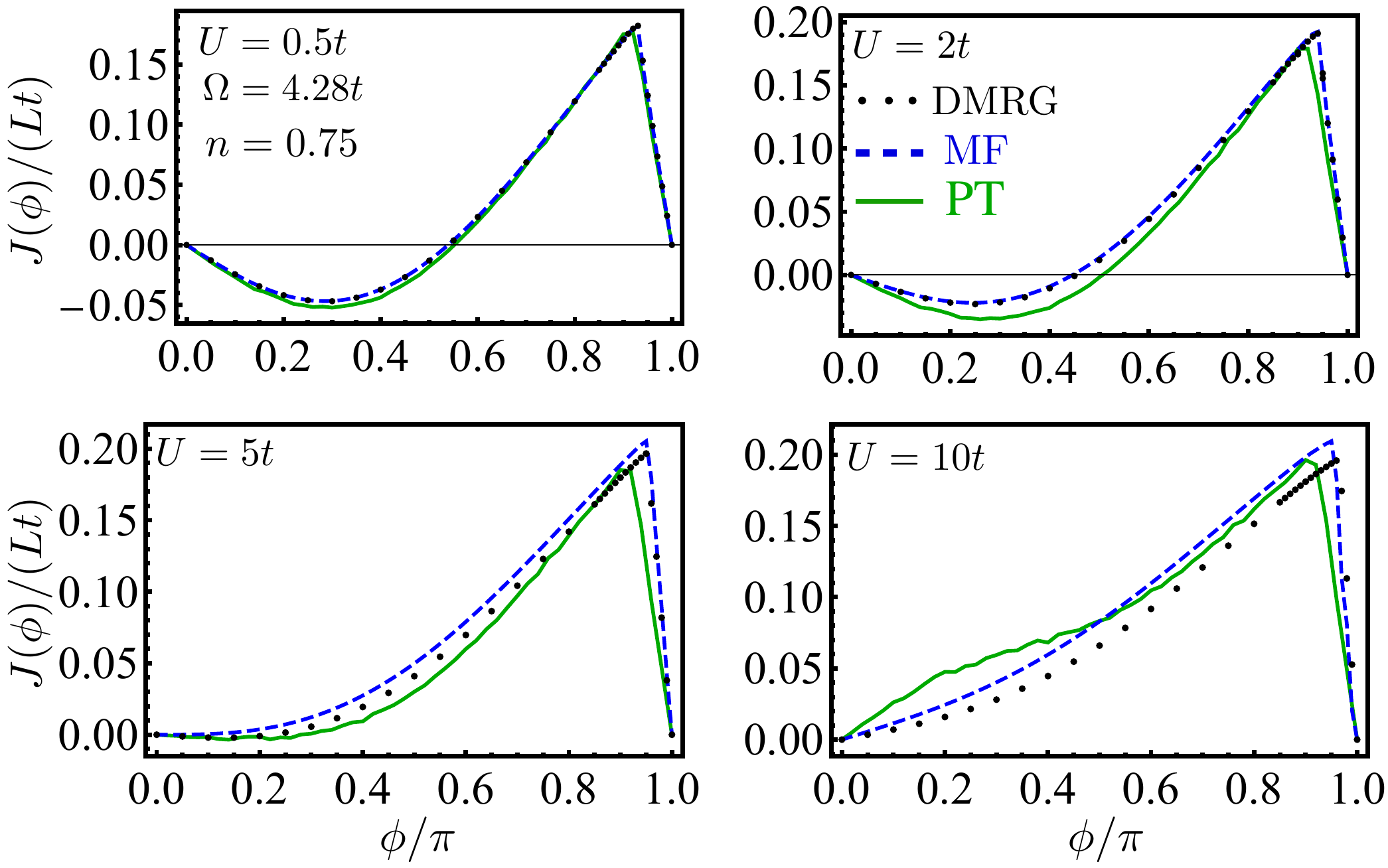}
\caption{ Comparison of different theoretical approaches to compute the current for different interaction strengths. The inter-leg hopping is $\Omega=4.28 t$, and filling $n=0.75$. The DMRG results is derived using system size $L=128$ and states $m=128$ which is well-converged. The DMRG results matches the mean-field result for interaction strengths up to $U\sim \Omega$.}\label{fig:comp}
\end{figure}
\begin{figure}[b]
\includegraphics[width=\columnwidth]{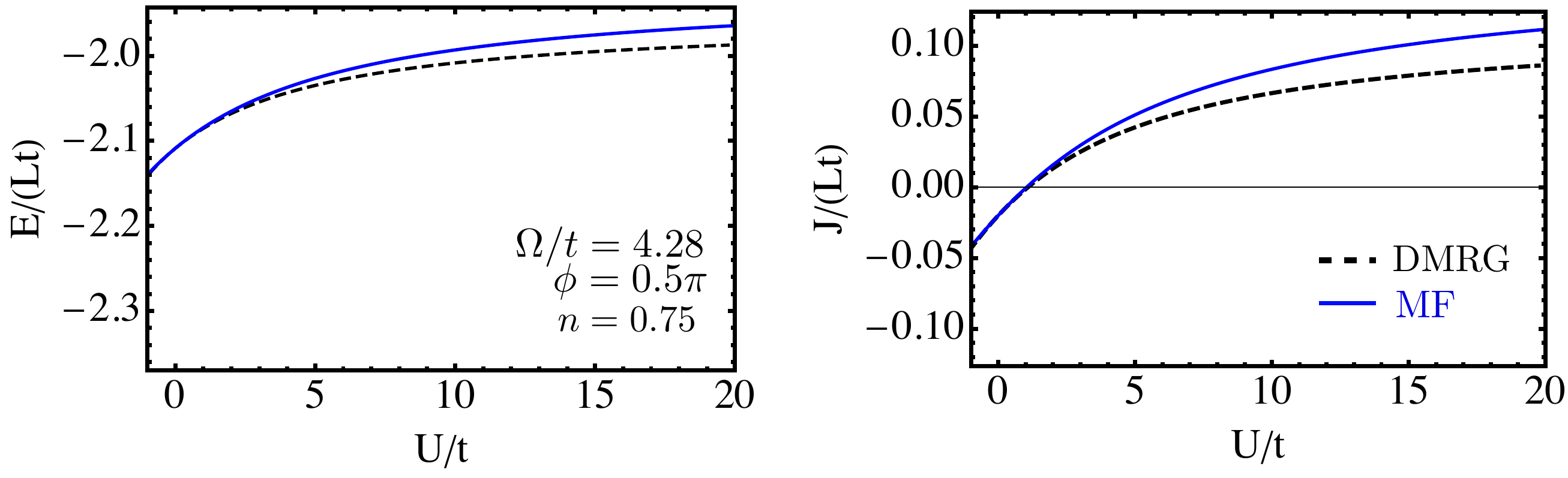}
\caption{
 Comparison of mean-field theory and DMRG: (a) Energy density as a function of the inter-leg interaction strength $U/t$ for $\Omega=4.28t$ and $\phi= \tfrac{\pi}{2}$ ($t$ is the intra-leg hopping). (b) Current density as a function of $U/t$. Mean-field theory is accurate up to $U\sim \Omega$. At saturation of the orbital current is observed at large $U$, which is not well captured by the present mean-field theory.}
 \label{fig:comp2}
\end{figure}

In the previous section, using essentially
numerically exact DMRG calculations, 
we have shown that the orbital current of the
interacting  system retains many features
of the non-interacting orbital current, in particular it exhibits a
cusp as $\phi$ approaches $\phi = \pi$. This result  
strongly suggests that  a conveniently 
 mean-field renormalized band structure should be able to  describe this behavior.
In this section,  such mean-field theory is 
developed for $U > 0$. 

 In order to apply  mean-field theory to the model Hamiltonian introduced in Sec.~\ref{sec:intro}, we define 
 the following fluctuation operators:
\begin{align}
\delta \hat{\alpha}_m&= \hat{\alpha}_m 
 - \alpha_m(\phi), 
 \\
\delta \hat{n}_{m,\sigma}&=  \hat{n}_{m,\sigma} - n_{m,\sigma}(\phi),
\end{align}
where $\hat{\alpha}_m = c^{\dagger}_{m,\uparrow}c_{m,\downarrow} $. Thus,  the order parameters
(see Appendix~\ref{app:MF} for details of their numerical evaluation in finite systems) are:
\begin{align}
\alpha_m(\phi) &= \langle \hat{\alpha}_m \rangle = \langle  c^{\dagger}_{m,\uparrow}c_{m,\downarrow} \rangle,\notag\\
 n_{m,\sigma}(\phi) &=  \langle \hat{n}_{m,\sigma} \rangle. \label{eq:mfparam}
 \end{align}
Note that the constraint of constant particle number requires that
\begin{equation}
\sum_{m} \left[ n_{m,\uparrow}(\phi) + n_{m,\downarrow}(\phi) \right] = N.
\end{equation}
In the mean-field approximation where the fluctuation energy is neglected 
(i.e. terms that are quadratic in the operators $\delta \hat{\alpha}(j)$ and $\delta n_{m\sigma}$ are thrown away), 
the interaction term becomes
\begin{align}
U&^{\text{MF}} =  U\sum_m\left[ -n_{m,\uparrow}n_{m,\downarrow}+n_{m, \uparrow}c_{m,\downarrow}^{\dag}c_{m,\downarrow}+n_{m,\downarrow}c_{m,\uparrow}^{\dag}c_{m,\uparrow}\right]\notag\\
&+ U\sum_m\left[ |\alpha_m(\phi)|^2 -\alpha_m(\phi) c_{m,\downarrow}^{\dagger}c_{m,\uparrow}-\alpha^{*}_m(\phi)c^{\dagger}_{m,\uparrow}c_{m,\downarrow}\right].
\end{align}
This expression allows us to write  the mean-field Hamiltonian as follows:
\begin{align}\label{eq:MF}
H^{\text{MF}}&=H_0+U^{\text{MF}}.
\end{align}
In order to be able to compare to the DMRG data, we have diagonalized the mean-field Hamiltonian \eqref{eq:MF} numerically in chains of length up to $128$ sites with open boundary conditions. The order parameters are  obtained self-consistently by solving the mean-field equations \eqref{eq:mfparam} (see Appendix~\ref{app:MF} for details). Using the self-consistent solution, we have obtained  the orbital current that is shown in Fig.~\ref{fig:comp}. Numerically, we find that $n_{m,\uparrow} = n_{m,\downarrow}$ and therefore the effect of the inter-leg interaction is to enhance the inter-leg hopping by renormalizing it to larger (average) value: $\Omega'(\phi)=2U\alpha(\phi) +\Omega$. The mean-field enhancement of the inter-leg hopping (cf. Fig.~\ref{fig:omegamft})  can be qualitatively understood by realizing that, for repulsive interactions,  configurations such like the one shown in Fig.~\ref{fig:mf} are  favored by both the inter-leg hoppeng ($\propto U$ in Eq.~\ref{eq:model}) and the interactions ($\propto U$ in Eq.~\ref{eq:model}). The renormalization of the inter-leg hopping also  qualitatively explains the enhancement of the orbital current, as a larger effective $\Omega$ makes it easier for the fermions to circulate around the plaquettes.

\begin{figure}[ht]
 \centering
\includegraphics[width=\columnwidth]{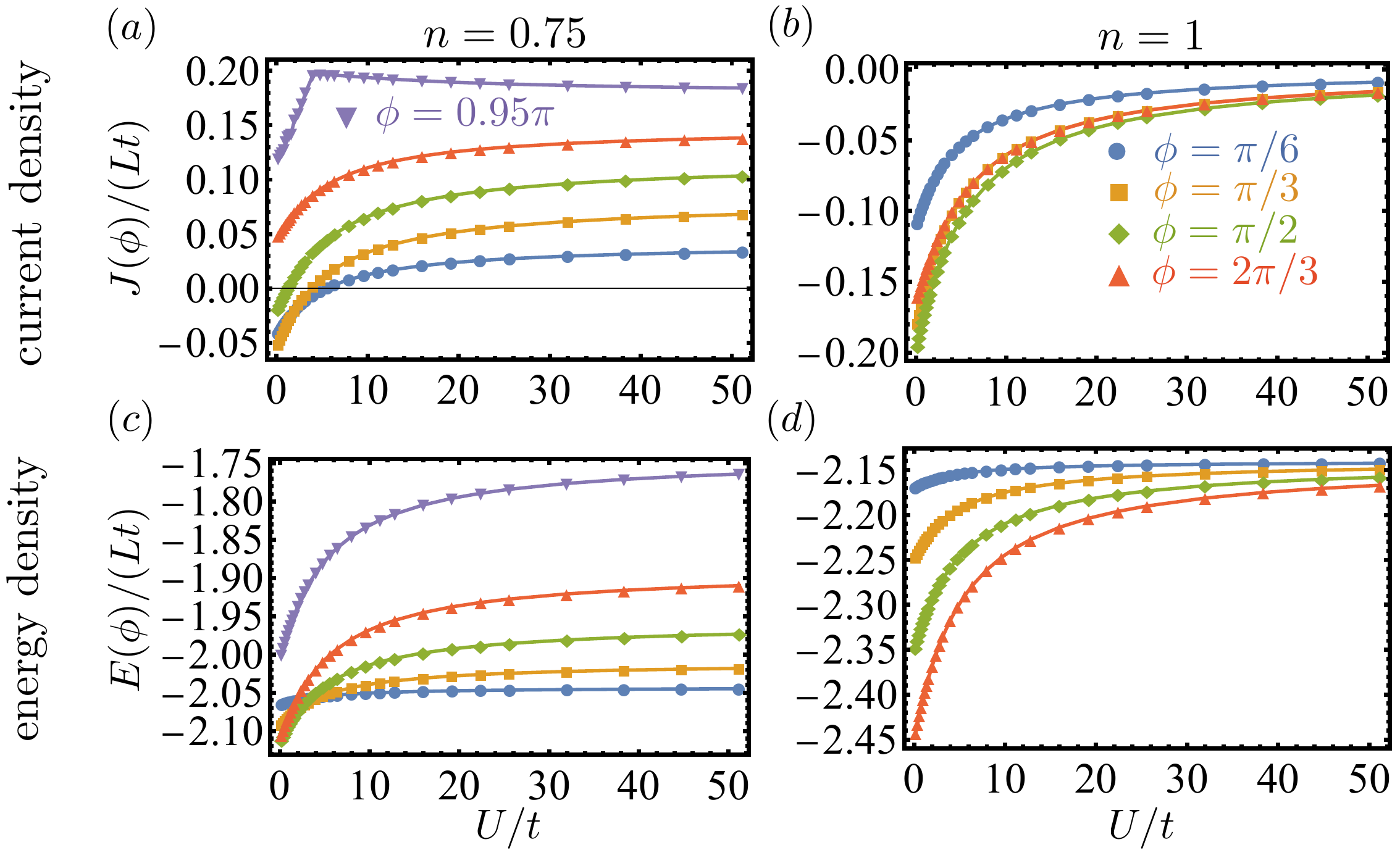}
\caption{Interaction-induced topological Lifshitz transition from DMRG. (a) Dependence of the current density on the interaction strength $U$ for a metallic lattice filling of $n=0.75$. Note there is a cusp in orbital current for $\phi= 0.95\pi$, which can be understood as an interaction-induced topological Lifshitz transition from a four Fermi-point to two Fermi-point metal with increasing $U/t$ (see Sect.~\ref{sec:strongU} and Fig.~\ref{fig:jp}). (b) Dependence of the energy density on $U$ for $n=0.75$. Note that both the energy and its derivative, i.e. the orbital current, appear to saturate at large $U$.
(c) Dependence of the current density on $U$ for the insulating filling $n=1$.(d) Dependence of the energy density on $U$ for $n=1$ filling. At this filling, the system is an insulator. Note the absence of a cusp for any value of $\phi$, which is different from the metallic case where $n=0.75$. The is no Lifshitz transition for this insulating filling}\label{fig:strongU}
\end{figure}

\begin{figure}[t]
\includegraphics[width=\columnwidth]{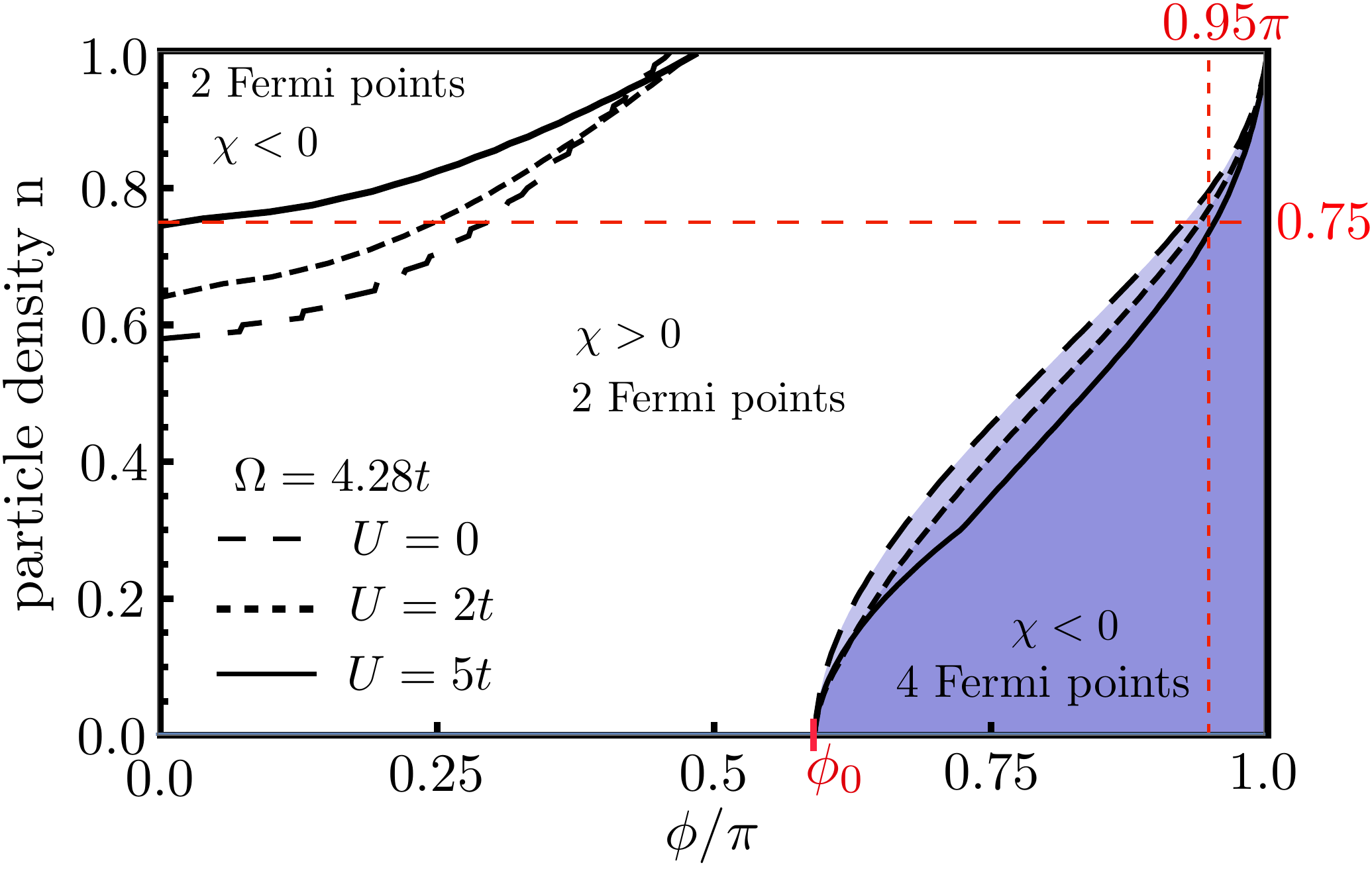}
\caption{Mean-field phase diagram for $\Omega=4.28t$ and different interaction strengths. The mean-field theory described in Sec.~\ref{sec:mft}  is accurate for $U\lesssim \Omega$. The dashed line indicates the lattice filling $n=0.75$ used in the upper panel in fig~\ref{fig:comp}. The interaction $U$ enhances the paramagnetic behavior in the two Fermi-point metallic phase, which  is overall rendered more stable by interactions. In this regard, note the mean-field theory predicts that  for $n=0.75$ and $\phi=0.95\pi$ the system undergoes a Lifshitz phase transition driven by the interaction at $U \simeq 4 t$, which is confirmed by our numerically exact DMRG calculations (see  Fig.~\ref{fig:strongU} and also Fig.~\ref{fig:jp}).  The value $\phi$ at which `double well' bands appear, $\phi_0$, is also indicated, see also Fig.~\ref{ladder}.}\label{fig:phase}
\end{figure}

Fig.~\ref{fig:comp}  shows the comparison  between the mean-field theory, perturbation theory (see Appendix~\ref{app:PT}), and the DMRG results for $n=0.75, \Omega=4.28t$ as a function of $\phi/\pi$ (Fig.~\ref{fig:comp}). Note that  the leading order perturbation theory is accurate as long as $4Ut^2\sin^2(\phi/2)/\Omega^3\ll1$ and over the considered parameter range is the less accurate of the three approaches compared in Fig.~\ref{fig:comp}.  Mean-field theory is most accurate for $U/\Omega \lesssim 1$. Fig.~\ref{fig:comp2} shows the comparison of the energy and current obtained within the mean-field theory and DMRG for $n=0.75,\Omega=4.28,\phi=\pi/2$ as a function of $U/t$.  In the large $U$ limit, the energy density and orbital current appear to saturate with increasing $U/t$. The limiting values at large $U$ are not fully captured by the mean-field theory, which is due to  correlation effects that are neglected  within the approach (see also Sec.~\ref{sec:correlation} below). Thus, for a completely accurate calculation of the orbital current in the regime $U\gg\Omega = 4.28 t$, we must entirely rely on DMRG.

   Based on the numerical observation that mean-field theory   accurately reproduces  both the energy and the orbital current  for $U \lesssim \Omega$,  we have obtained the mean-field phase diagram shown  in Fig.~\ref{fig:phase}. The boundaries of the phase diagram are calculated by locating the changes of sign and singularities of the orbital susceptibility $\chi(\phi,n)$, which can  be also obtained from  the mean-field theory. Thus,  we find a decrease in size of the diamagnetic region as the strength of $U$ is increased (cf. Fig.~\ref{fig:phase}). The behavior at zero field can be also understood  by using the mean-field renormalized inter-leg hopping $\Omega'(\phi)$ in Eq.~\eqref{eq:chi0} from the Appendix.

   In addition, in Fig.~\ref{fig:phase}, we have used a dashed line to indicate the phases available for $n=0.75$ and $\phi=0.95\pi$ for different values of $U$, so that comparison with the upper-left panels of Figs.~\ref{fig:strongU} and ~\ref{fig:jp} is  possible. Notice that,  for this lattice filling and flux,  mean-field theory predicts  an interaction-driven Lifshitz transition between four and two Fermi points phases. This result is consistent with the cusps in the orbital current observed in the  DMRG results on Fig.~\ref{fig:strongU} [panel (a)], which shows  a cusp in the orbital current   for $U/t \simeq 4$ (see also following section).   On Fig. ~\ref{fig:strongU} we also show the behavior of the orbital current for $n=1$, for  which the system is a fully gapped ferromagnetic insulator. In this phase, the orbital current changes smoothly with increasing $U/t$ and $\phi$ and no cusps in the current are observed as no Lifshitz transitions can occur. 


\subsection{Interaction-induced Lifshitz transition}\label{sec:strongU}

Finally, let us briefly discuss  the limit of $U\gg t,\Omega$ close to unit lattice filling.  In this parameter regime, the intra-chain hopping  $H_{\parallel}$ can be treated as a the smallest term in the Hamiltonian. Second order perturbation theory at unit filling  allows to map the Hamiltonian onto a   spin-chain model with Dzyaloshinskii--Moriya (DM) interactions in a transverse magnetic field $\propto \Omega$~\cite{PhysRevB.73.195114}. For $\Omega = 0, \phi=0$, the system is a Tomonaga--Luttinger liquid with charge gap but no spin gap~\cite{giamarchi2003}.   For $\phi,\Omega\neq 0$, as obtained in Ref.~\cite{PhysRevB.88.165101} (in the uncoupled chain limit corresponding to $J_{\perp} = 0$ in the notation of that article),  the spin chain  undergoes a commensurate-incommensurate phase transition from the   spin-gapless Tomonaga--Luttinger  to a fully gapped ferromagnetic insulator~\cite{PhysRevB.73.195114}. Although  to the
best of our knowledge, there is no analytical expression for the critical curve $\Omega_c(\phi)$),   $\Omega_c\lesssim 4t^2/U$  for all $\phi$ being $\Omega_c = 0$ for $\phi = \pi/2$~\cite{PhysRevB.88.165101}. Furthermore, since the map to a spin chain  requires that $U \gg t$, $\Omega_c$ is therefore small and lies outside the large $\Omega$ regime studied in this article. 

 Nevertheless, in the large $U$ and $\Omega$ limit, it is tempting to regard  the system at $n \lesssim 1$ as a ferromagnetic   metal   resulting from doping the ferromagnetic insulator described in the small $\Omega$ limit in Ref.~\cite{PhysRevB.88.165101}.   In  mean-field theory, this ferromagnetic  metal has the band structure described in Sec.~\ref{sec:mft}, and thus exhibits a large spin gap of the order of the renormalized inter-leg hopping   $\Omega^{\prime}(\phi) \sim \Omega$. Depending on  $n$ and $\phi$, this band structure supports two or four Fermi points in the lower band for $n < 1$ and therefore a topological Lifshitz transition between these two  phases is possible.   If correlation effects beyond single-particle (mean-field) picture are further accounted for, the system is described as a Tomonaga--Luttinger liquid with zero charge gap and a finite and large spin gap. This conclusion was also reached in Sec~\ref{sec:largeomega}. Indeed, we believe this is a sensible picture for the two-Fermi point metallic phase except possibly at half-filling (i.e. $n=1/2$), for which umklapp scattering could drive a metal-insulator transition~\cite{giamarchi2003} and open a charge gap (whether this is the case or not requires further analysis, which will be presented elsewhere~\cite{unpub}).

 However, in the presence of interactions, it is not clear whether the four Fermi point metallic phase remains fully gapless as suggested by mean-field theory.   This is because the doubling in the number of Fermi points from two to four implies
 a doubling  of the number of gapless low-energy degrees of freedom. In the presence of interactions and for generic fillings, a subset of the low energy degrees of freedom may become gapped by certain scattering processes~\cite{giamarchi2003}. To see how this could happen, let us use  the following continuum-limit representation of the  fermion operator for the lower (partially-filled) band with four Fermi points:
\begin{align}
c_{m,\sigma}&\sim A_{R1,\sigma} \psi_{R1,\sigma}(x_m) e^{-i k_{F1} x_m} \notag\\
&\quad + A_{R2,\sigma} \psi_{R2,\sigma}(x_m) e^{+i k_{F2} x_m}\notag\\
&\quad + A_{L1,\sigma} \psi_{L1,\sigma}(x_m)e^{+i k_{F1} x_m}\notag \\
&\quad + A_{L2,\sigma}\psi_{L2,\sigma}(x_m)e^{-i k_{F2} x_m}, \label{eq:fmop}
\end{align}
where  $x_m = m$ and $\psi_{R/L, 1/2}(x)$ are Fermi fields that vary slowly on the scale of the lattice and describe the
low-energy degrees of freedom of the system~\cite{giamarchi2003}. The coefficients $A_{R1/2,\sigma}, A_{L1/2,\sigma}$
are related to the angle $\theta(q,\phi)$ introduced when diagonalizing the kinetic energy (see Eq.~\ref{eq:thetaq} 
in Appendix~\ref{sec:band}). Thus,  we find that the continuum limit of the intra-leg interaction, Eq.~\eqref{eq:int} contains the following two types of scattering processes between the two sets of Fermi points in the lower band:
\begin{align}
H_{U} &= H_{+} + H_{-} +  \cdots  \label{eq:cont}\\
 H_{+}  &=  g_{+} \int dx \,  \psi^{\dag}_{R1}(x)  \psi^{\dag}_{L1}(x)   \psi_{L2}(x)    \psi_{R2}(x)  + \mathrm{H.c.}\\
 H_{-}  &=  g_{-} \int dx \,  \psi^{\dag}_{L1}(x)  \psi^{\dag}_{R1}(x)   \psi_{L2}(x)    \psi_{R2}(x)  + \mathrm{H.c.}
\end{align}
The ellipsis in  Eq.~\eqref{eq:cont} stands for forward scattering and other terms that oscillate rapidly on the  lattice scale. The scattering processes  described by $H_{\pm}$ correspond to processes for which a fermion is scattered
by another fermion
 from one Fermi point to another by exchanging a momentum $k_{F2}\pm k_{F1}$ 
 (recall that $k_{F2}-k_{F2} = \pi n$ but the individual values of $k_{F1,2}$ depend
on the band dispersion). These scattering processes can potentially become relevant
perturbations in the renormalization-group (RG) sense, which would result in the gaping some of the
low-energy degrees of freedom of the four Fermi-point metal. However, note that the scattering process described by $H_{-}$ is obtained by exchanging the final states in the process described by $H_{-}$. Hence, the relation $H_{-} = - g_{-} H_{+}/g_{+}$ follows. In addition, the coupling constants $g_{\pm} = A_{R1} A_{R2} A_{L1} A_{L2}  U$ in the small $U$ limit and therefore, for the type of interactions considered here, $H_{-}$  appears to cancel $H_{+}$ 
in the weak coupling limit.   A  more detailed perturbative analysis for  general interactions of the
flow of the $g_{\pm}$ couplings under RG scaling is beyond the scope of this work and will be presented elsewhere~\cite{unpub}.

 We have also attempted to numerically address the issue of the existence of a correlation gap in the four Fermi-point phase at finite $U$ by  computing the central charge of the system from the scaling of the subsystem entanglement entropy, which is one of the outputs of the DMRG code. 
We have found strong evidence that the central charge for the phase
across the Lifshitz transition shown in Fig.~\ref{fig:strongU} (a) for $U/t = 5$, $\phi = 0.95\: \pi$, and $n=0.75$ 
is $c=1$ as corresponds to a two Fermi-point metallic phase. However, 
the situation is less clear for $U/t \lesssim 4$, for which $c = 2$ is expected. This is because the entanglement entropy shows rather pronounced oscillations~\cite{Sorensen_PhysRevB.87.115132}
as a function of  the subsystem size.  Even after extending our calculations  
up to $L = 512$ sites and $m = 512$ states, we were unable to reach a definite conclusion
on the central charge of the metallic phase  at small $U/t$ for $\phi=0.95 \: \pi$
and $n=0.75$. However, let us point out that the DMRG results  discussed in the following section for the Friedel oscillations of the inter-leg current  appear to be consistent with the absence of a interaction-induced gap.

%
\subsection{Friedel oscillations of the Inter-leg current}\label{sec:correlation}
\begin{figure}[t]
\includegraphics[width=\columnwidth]{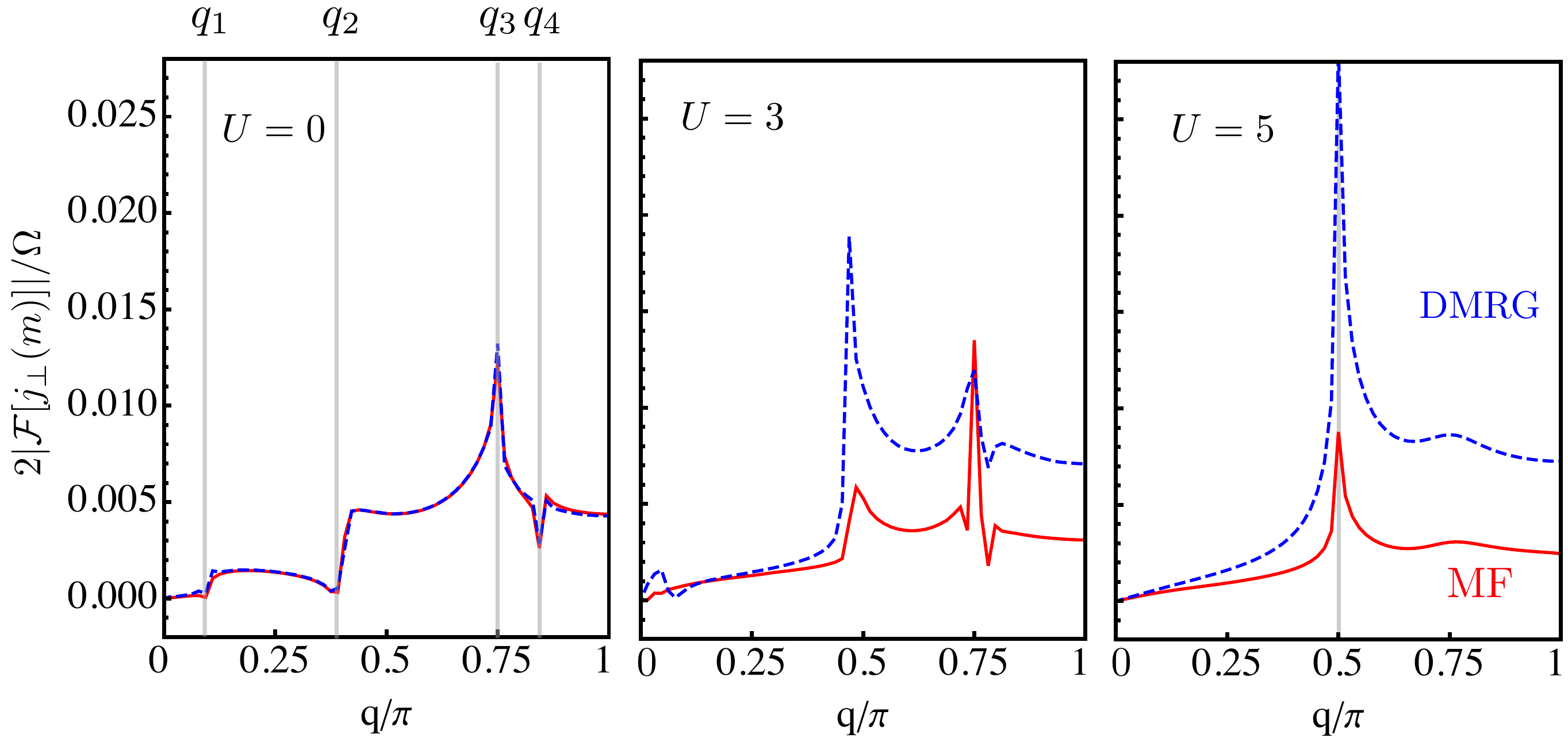}
\caption{(a)  Absolute value of the Fourier components of the inter-leg current $j_{\perp}(m)$ for particle filling $n=0.75$ and $\phi=0.95\pi$ computed using DMRG (dashed line) and exact diagonalization at $U=0$ (continuous line). The curves exhibit peaks located at wave numbers  $q_1=
k_{F2}-k_{F1}$, $q_2=k_{F2}+k_{F1}$, $q_3=2k_{F2}$, and $q_4=2k_{F1}$ where $\pm k_{F1}$ and $\pm k_{F2}$ are the Fermi momenta in the four Fermi-point phase. (b,c) Absolute value of Fourier components of the inter-leg current for inter-leg interaction strength $U=3t, 5t$. Unlike the non-interacting case, the intensity of the spectrum obtained from mean-field theory differs from the DMRG result, which is due to correlation effects. However, notice that the position of the peaks is the same for mean-field theory and DMRG. Panel (c) shows the system is in a metallic phase with only two Fermi points and therefore the spectrum shows a single peak at $q = (2\pi -) 2k_F$. The Fourier spectrum in all panels is $2\pi$ periodic and symmetric with respect to $q=0$ with the Fourier components at $q=0$ vanishing by Kirchoff's law. The peaks at wave numbers $q_1,q_2,q_3,q_4$ must be folded within the first Brioullin zone which is $q=(-\pi,+\pi]$.}\label{fig:jp}
\end{figure}

So far, we have discussed the behavior of the average orbital current density using DMRG,  mean-field theory, and perturbation theory.  As described in the previous section, mean-field theory deviates from  DMRG  in the strongly interacting limit (cf. Fig.~\ref{fig:comp}), where it fails to fully capture saturation of both the energy and current densities.  Besides this, as we show below,  the decay of the oscillations of the inter-leg orbital current is also not  correctly reproduced by the mean-field approach for finite $U$. The inter-leg current is defined by the (ground-state expectation value of the)  imaginary part  of the inter-leg hopping operator, i.e.
\begin{align}
j_{\perp}(m)=-\frac{\Omega}{2}\text{Im}\left[\left\langle\hat{\alpha}_m\right\rangle\right] = -\frac{\Omega}{2}\text{Im}\left[\left\langle c^{\dag}_{m,\uparrow} c_{m,\downarrow} \right\rangle\right]. \label{eq:jperp}
\end{align}
In lattices with periodic boundary conditions,  translational invariance implies that $j_{\perp}(m)$ is a constant independent of $m$, which, by Kirchhoff's laws,   vanishes. However, because in our mean-field and DMRG calculations are carried out in systems with open boundary conditions, $j_{\perp}(m)$ displays an oscillatory behavior as we describe below.

 In order to characterize the oscillatory behavior of the inter-leg current, we compute its Fourier transform defined as:
\begin{equation}
\mathcal{F} \left[j_{\perp}\right](q) = \frac{1}{\sqrt{L}}\sum_{m=1}^L e^{i q m} j_{\perp}(m).
 \end{equation}
In Fig.~\ref{fig:jp}, we have plotted the absolute value of the Fourier components of the inter-leg current obtained from DMRG and mean-field theory for $n=0.75$ and $\phi=0.95\pi$. We notice that for $U/t= 0$ and $U/t = 3$ (left panel),  the spectrum displays singularities   at a set of wave numbers that correspond to $q_1=k_{F1}+k_{F2}, q_2= k_{F2}  - k_{F1}, q_3=2k_{F2}, q_4=2k_{F1}$ (actually at the corresponding values folded into the positive 1BZ, because  $\mathcal{F} \left[j_{\perp}\right](q)$  is periodic with periodicity $2\pi$). These singularities in the Fourier spectrum of $j_{\perp}$ are a consequence of the
reflection of the fermions at the boundaries, which leads  to Friedel oscillations. 
The wave numbers
of the oscillations can be obtained  by introducing the low-energy expansion of the fermion operator around the four Fermi points given in Eq.~\eqref{eq:fmop} into Eq.~\eqref{eq:jperp}.
The behavior of the Fourier components of the inter-leg current close to these set of wave numbers is 
related to decay with  $m$ (away from the left boundary, or $L - m$ away from the right boundary) 
of the different oscillation harmonics  of $j_{\perp}(m)$~\cite{Cazalilla_2004,PhysRevB.76.195105}. 
 On the other hand, the rightmost panel of  Fig.~\ref{fig:jp} shows the Fourier spectrum of $j_{\perp}(m)$ computed at $U/t = 5$ from DMRG and  mean-field theory. In this case, the singularities  are located at $2\pi -2 k_F = 0.5 \pi$  (recall that $q = 2k_F= 2 \pi n = 1.5 \pi$), as expected for a metallic system with  two Fermi points at $\pm k_F$. These results provide  numerical evidence that allows to relate the cusp  in the orbital current shown in Fig.~\ref{fig:strongU} for $\phi/\pi = 0.95,\, n=0.75$ to an interaction-induced Lifshitz transition from the four to the two-Fermi point phase, as predicted by the mean-field theory.   

  The central and rightmost panels of Fig.~\ref{fig:jp} also show that the mean-field theory fails to reproduce the magnitude of Fourier spectrum of the inter-leg current computed using DMRG results.  Indeed, the numerically exact DMRG results display stronger singularities  than predicted by the mean-field theory. The latter approach treats particles as independent (i.e. uncorrelated) fermions that move in an effective potential, giving rise to the renormalized inter-leg hopping $\Omega^{\prime}(\phi)$. In this regard, the (boundary) exponents of the power laws that control the decay of the oscillations away from  $m= 0, L$ are the same as for an non-interacting system.  However, as we have mentioned above, the system is a spin-polarized Tomonaga--Luttinger liquid and therefore the boundary exponents  are  interaction dependent and different from the non-interacting exponents~\cite{giamarchi2003,Cazalilla_2004,PhysRevB.76.195105}.  The singularities displayed by the Fourier transform of the inter-leg current in our calculations  are
affected by finite-size effects, which complicates the extraction of such boundary exponents.  However, qualitatively as simple eye inspection shows, we observe the peaks in $\mathcal{F}[j_{\perp}](q)$  are less singular than their mean-field (i.e. non-interacting) counterparts.  The presence of these singularities is consistent with the system being gapless on both sides of the interaction-induced Lifshitz transition. Or, if a interaction-induced gap exists in the four Fermi-point phase for $U/t \lesssim 4$, the latter remains small for the system sizes of $L = 128$ studied here. As mentioned above,  additional studies of this issue will be reported elsewhere~\cite{unpub}.


\section{Summary, discussion, and outlook}\label{sec:conclu}

In this work, we have investigated the behavior of the orbital current in a model of a two-leg ladder  relevant to
ultra-cold Fermi gases subject to synthetic gauge fields. We have discussed how the orbital current, which is a gauge invariant observable that can be measured in electronic systems, can be also 
measured in these atomic setups. In addition, we have seen that, at zero temperature, the orbital current exhibits cusps either as a  function of the (pseudo-) gauge field flux and the interaction. In the  non-interacting limit, the cusps 
can be related to topological Lifshitz transitions, where the number of Fermi points changes. 
Furthermore, in the non-interacting limit, 
we have also investigated the analytical form of the singularities in the orbital current and the orbital 
susceptibility, the latter being the derivative of the orbital current with respect to the flux. Thus, we have found different types of
singular behavior depending on whether the particle number or the chemical potential
are fixed. 

Interestingly,  in the presence of  repulsive interactions, we have also observed cusps in the orbital current up to values of the interaction strength $U$ much larger than the inter-leg hopping $\Omega$, which, as in the experiments~\cite{Mancini1510_sun,PhysRevLett.117.220401_sun},   we have taken to be larger than the intra-leg hopping.  For intermediate  interactions,
i.e. for $U \lesssim \Omega$, we have shown that mean-field theory is capable of accounting for the cusps in the orbital current. This has allowed us to  obtain a mean-field phase diagram by monitoring the changes of sign and singularities
of the orbital susceptibility. At small flux per plaquette, we have found that interactions can
change the sign of the susceptibility and thus change the behavior of the system from paramagnetic to
diamagnetic. At large flux per plaquette, the mean-field theory predicts an interaction induced Lifshitz
transition at which the number of Fermi points of the renormalized (mean-field) band structure 
changes from four to two at fixed particle number. The results obtained
from unbiased DMRG  calculations of  the Friedel oscillations of the inter-leg current further 
support this prediction of mean-field theory.

It is interesting to note that, despite  the orbital current being a non-universal quantity that is 
contributed by all fermions~\cite{PhysRevB.76.195105,PhysRevB.73.195114}
 and not only by those in the vicinity of the Fermi points,  its behavior is very sensitive to the topological
 properties of the Fermi surface (in the 1D case, the number of disconnected components) . However, 
 what may be most striking of the signatures of Lifshitz transitions discussed
 above is their robustness to the presence of strong interactions. This is perhaps most surprising 
 in the case of  one dimensional  systems studied here, for which  the concept of a
  ``band structure'' is often hardly justified as interactions completely wash out  
  Landau quasi-particles. The latter are  ubiquitous
 in two and three dimensional Fermi systems~\cite{giamarchi2003,baym2008landau}, and this 
 provides a theoretical justification for the use of mean-field theory in the calculation of band structures
 and other spectral properties of two- and three-dimensional systems. It would be
 interesting to extend the type of studies reported here to higher dimensional systems. If our 
 expectations are correct, the  existence of stable Landau quasi-particles should make the predictions 
 of  mean-field theory for the orbital current even more reliable. The difficulty in this case may be to find an unbiased 
 method like DMRG  to test the mean-field predictions against accurate numerical calculations of the orbital current
 and susceptibility.  

 As far as one dimensional systems are concerned, this study has raised some issues that need to be addressed in more depth in future studies. One of them has been already pointed above, and concerns the investigation of interaction-induced gaps in the low energy sector and their impact on the behavior of the  the orbital current and susceptibility close to a Lifshitz transition. This may be especially relevant for  systems systems with attractive interactions (e.g. for $U < 0$) as in this case the interaction and the inter-leg hopping favor different  types of configurations in the ground state, therefore compete with each other. For repulsive
 interactions a careful investigation of the system at half-filling (i.e. for $n=\tfrac{1}{2}$) is also  necessary, as it has also been pointed out above.  In addition, more accurate numerical
 studies using DMRG are required in order to understand the behavior of the 
 entanglement entropy across the Lifshitz transition, and to assess to what
 extent the singularities exhibited by the orbital susceptibility are modified by the
 interaction effects. The latter study will require a more careful finite-size analysis 
 of the orbital current and susceptibility close to the transition.

Finally, in connection with recent~\cite{Mancini1510_sun,PhysRevLett.117.220401_sun} and future experiments, finite temperature effects should be also considered. The extension of the mean-field theory developed here to finite temperatures is rather straightforward, but unbiased calculations of the orbital current or susceptibility at finite temperature using DMRG
will be more demanding. Qualitatively, we expect that the quantum phase transitions of the Lifshitz
type studied here will become crossovers, and finite temperature effects will round  the sharpness
 of cusps in the orbital current and the singularities of the susceptibility.  


\section*{Acknowledgments}

The work of M.T. was partly supported by Grants-in-Aid No.~JP17K17822, No.~JP20H05270, No.~JP20K03787, and No.~JP21H05185 from MEXT of Japan. The work of MAC has been supported by Ikerbasque, Basque Foundation for Science and  MICINN grant no. PID2020-120614GB-I00 (ENACT).

\appendix

\section{Bloch-wave Basis} \label{sec:band}

 In this Appendix, we review the band structure of the system described by~\eqref{eq:model} in the non-interacting limit (i.e. for $U=0$). In order to obtain the band dispersion, we follow~\cite{PhysRevB.76.195105} and  
applying a Fourier transform to the kinetic energy in Eq.~\eqref{eq:model},
\begin{align}
 c_{q,\sigma}&=\frac{1}{\sqrt{L}}\sum_{m}e^{i(q+\phi_{\sigma})m}\tilde{c}_{m,\sigma},
\end{align}
followed by a unitary transformation in  the ladder indices $\sigma$. In the above equtation $\phi_{\sigma}$ is gauge phase for species (leg) $\sigma$.
Setting $\phi_{\downarrow}-\phi_{\uparrow}=\phi$,  the flux-dependence from the inter-leg hopping is eliminated, which
renders the Hamiltonian to the following form:
\begin{multline}
\hat{H}_0=\sum_{q\in \mathrm{1BZ},\sigma} \left[-2t\cos \left(q+\phi_\sigma\right)\:  c^{\dagger}_{q,\sigma} c_{q,\sigma} \right] \\ - \frac{\Omega}{2} \sum_{q\in \mathrm{1BZ}} \left( c^{\dagger}_{q,\uparrow} c_{q,\downarrow}+   c^{\dag}_{q,\downarrow}  c_{q,\uparrow} \right),
\end{multline}
where $1\mathrm{BZ}$ corresponds to the segment $(-\pi,\pi ]$ of lattice  momenta in units where the lattice parameter $a =1$. The above Hamiltonian can be diagonalized by means of the following unitary transformation:
\begin{equation}\label{eq:tr}
\begin{pmatrix} 
 c_{q,\uparrow}\\
 c_{q,\downarrow}
\end{pmatrix}
=
\begin{pmatrix}
\cos\left[\theta(q,\phi)\right]&\sin \left[\theta(q,\phi)\right]\\
-\sin\left[\theta(q,\phi)\right]&\cos\left[\theta(q,\phi)\right]\\
\end{pmatrix}
\begin{pmatrix} 
c_{q,u}\\
c_{q,d}
\end{pmatrix}.
\end{equation}
Hence, 
\begin{equation}
\hat{H}_0=\sum_{q \in \mathrm{1BZ}} \left[  \epsilon_u(q,\phi) \: c_{q,u}^{\dagger}c_{q,u}+\epsilon_d(q,\phi)\:  c^{\dagger}_{q,d}c_{q,d} \right],
\end{equation}
where the single-particle dispersion $\epsilon_{s} (q,\phi)$  is given by
\begin{align}
\epsilon_{s} (q,\phi) &=-2t \cos\left(q+\frac{\phi_{\uparrow}+\phi_{\downarrow}}{2}\right)\cos(\phi/2)\notag\\
&\pm \frac{1}{2}\sqrt{\Omega^2+\left[4t\sin\left(q+\frac{\phi_{\uparrow}+\phi_{\downarrow}}{2}\right)\sin(\phi/2)\right]^2}.
\end{align}
The rotation angle of the unitary transformation can be obtained from the equation:
\begin{align}
\sin&[\theta(q,\phi)]^2=\frac{1}{2}\left[1-\frac{4 t\sin(q+\frac{\phi_{\uparrow}+\phi_{\downarrow}}{2})
\sin \frac{\phi}{2}}{\epsilon_{u}(q,\phi)-\epsilon_{d}(q,\phi)}\right], \label{eq:thetaq}
\end{align}
and $s = u, d$, where $u \to +$ ($d\to -$) stands for upper (lower) band. In Fig.~\ref{ladder}, we show the band structure as a function of the flux $\phi$ per plaquette and the ratio of hopping amplitudes,  $\Omega/t$. The red vertical line corresponds to the experimental situation considered in Ref.~\onlinecite{PhysRevLett.117.220401_sun}, which is in the large $\Omega/t$ regime. The band structure can be classified into four types according to whether there is band overlap between the upper and lower bands (cases A and B) or not (cases C and D), and whether the curvature of the lower band at $k=0$ is positive (A and D) or negative  (B and C, cf. Fig.~\ref{ladder}).  Note that, whenever there is no band overlap as it is the case for sufficiently large $\Omega/t$
the lower band curvature  determines whether only two Fermi points can exist (positive curvature at $q = 0$) or four Fermi points can exist for certain band filling (negative curvature at $q=0$). 

Next, we briefly consider the effects of the interaction in the basis of Bloch states that diagonalize the kinetic energy.  In this basis, the inter-species (inter-leg) interaction takes the form:
\begin{equation}
\hat{U}= \frac{1}{L}\sum_{pkq} \tilde{U}(p,k,q;\phi) c_{p,d}^{\dagger}c_{k,d}^{\dagger}c_{k-q,d}c_{p+q,d}  + \hat{U}^{\prime}, \label{eq:intu}
\end{equation}
where $U^{\prime}$ involves the creation and destruction operators in the upper band. In the  above expression:
\begin{align}
\tilde{U}(p,k,q;\phi)&=U\Gamma(p,k,\phi)\Gamma(p+q,k-q,\phi),\\
\Gamma(p,k,\phi)&=\frac{1}{2}\left\{ \sin\left[\theta(p,\phi)\right]\cos
\left[\theta(k,\phi)\right]-\sin\left[\theta(k,\phi)\right]\notag \right.\\
&\left. \times\cos\left[\theta(p,\phi)\right]\right\}.
\end{align}
In  the regime of interest in this work where $\Omega \gtrsim 4 t $, there  is a large gap $\sim \Omega$ between the lower and upper bands. Therefore, at low temperatures and for weak interactions (i.e. $U\ll\Omega$) and low filling ($n \leq 1$),  it is safe to neglect scattering processes involving states in the upper band described by $U^{\prime}$ in Eq.~\eqref{eq:intu} . In this regard, we note that, for small $\phi$, $\Gamma(p,k,\phi= 0)\sim \phi^2$, which means that, for $2t/\Omega\ll 1$ and $U\ll \Omega$, the (inter-leg) interaction has a rather weak effect on the system, as it has been already noticed in Sec.~\ref{sec:largeomega} by working  in a different basis. Indeed, in the limit where $2t/\Omega\ll1$, we have
\begin{align}
\tilde{U}&=U\left(\frac{2t}{\Omega}\right)^2 \sin^2(\phi/2) \cos^2(Q)\sin(q_1)\sin(q_2)\notag\\
&+O\left[\left(\frac{2t}{\Omega}\right)^4\right],
\end{align}
where $Q=(p+k)/2$, and $q_1=(k-p)/2, q_2=(k-p-2q)/2$. This expression makes it clear that the 
effect of the interactions is suppressed by powers  of $(t/\Omega)^2 \ll 1$ in the large $\Omega$ limit.
Note as well that for $U\gtrsim \Omega$  and $\phi \sim \pi$, the effect of the interaction becomes more important. 

\subsection{Perturbation theory}\label{app:PT}

  In the weakly interacting case (i.e. for $4Ut^2\sin^2(\phi)/\Omega^3 \ll1$ as pointed out above), the chemical potential can be  approximately determined using non-interacting model since the chemical potential is of order $-\Omega/2$ and is much larger than the Hartree-Fock shift $\sim 4Ut^2\sin^2(\phi)/\Omega^2$. Thus, from Eq.~\eqref{eq:JJ}, the orbital current can be calculated from the effective Hamiltonian:
\begin{align}
J(N,\phi)
&=-\sum_{\sigma=\uparrow,\downarrow}\sum_{q}\frac{\partial \epsilon_{q,\sigma}(\phi)}{\partial \phi} n_{q,\sigma}(\phi)\notag\\
&-\sum_{p,k,q}\frac{\partial \tilde{U}(p,k,q;\phi)}{\partial\phi}\left\langle c_{d,p}^{\dagger}c_{d,k}^{\dagger}c_{d,k-q} c_{d,p+q}\right\rangle.\label{eq:Jp}
\end{align}
Hence,  the orbital susceptibility can be obtained as $\chi=L^{-1}\partial J/ \partial\phi$.

The leading order contribution to the orbital current  stems from the the interaction term since 
the first-order  correction to momentum distribution vanishes. This is nothing but the derivative of  the Hartree-Fock correction
to the total energy:
\begin{align}
J^{(1)}&=\frac{1}{L}\sum_{p,k,q}\frac{\partial \tilde{U}(p,k,q;\phi)}{\partial\phi}\left\langle c_{d,p}^{\dagger}c_{d,k}^{\dagger}c_{d,k-q}c_{d,p+q} \right\rangle_0\notag\\
&=\frac{2U}{L}\sum_{p,k}\frac{\partial \Gamma(p,k;\phi)^2}{\partial\phi}n_{d,p}n_{d,k},
\end{align}
where $n_{d,p}$ stands for the non-interacting Fermi-Dirac distribution. The total orbital current including the above correction is shown on the left panel of Fig.~\ref{fig:comp}. We note the leading order correction suffices to qualitatively capture the interaction-induced change from diamagnetic to  paramagnetic behavior at small $\phi$ in the regime where $4Ut^2 \sin^2(\phi/2)/\Omega^3\ll1$.



\section{Results for Non-Interacting Fermions}\label{sec:nonint}

\subsection{Orbital Current at fixed particle number}

In this Appendix, we review the most important results for the 
orbital current in the absence of interactions. As explained in Sec.~\ref{sec:eff}, 
the orbital current can be calculated from  partition function,  which for the non-interacting
system reads:
\begin{align}
Z_0=\prod_{s=u,d}\prod_{q}\left[1+e^{-\beta\epsilon_{s,q}(\phi)+\beta \mu(\phi)}\right].
\end{align}
Hence, the free energy for fixed particle number is obtained:
\begin{align}
G^{(0)}(N,\phi)=-\frac{1}{\beta}\log(Z_0)+N\mu(\phi).
\end{align}
Thus, the orbital current is:
\begin{align}
J^{(0)}(N,\phi)=-\frac{\partial}{\partial \phi}G^{(0)}(N,\phi)
=-\sum_{s=u,d}\sum_{q}n^{(0)}_{s,q}\frac{\partial \epsilon_{s,q}(\phi)}{\partial \phi},
\end{align}
where $n^{(0)}_{s,q}$ is the Fermi distribution function. Fig.~\ref{fig:j0} shows the results of evaluating this expression for the orbital current density $J^{(0)}(N,\phi)/L$ at zero temperature for several values of the lattice filling $n= N/L$. We note that there are two non-trivial features in the behavior of current: (1) At large lattice filling and small $\phi$, the sign of the orbital current is the opposite (negative in our convention) to  the sign at low
lattice fillings.  And (2) Near $\phi=\pi$, there is a cusp, which is related to the number Fermi points in the band structure transitioning from two to four.  As discussed in the main text, the cusp is related to  a change in the number of Fermi points, which in one dimension is an example of a topological Lifshitz transition. 


\subsection{Orbital Susceptibility near $\phi=0$}\label{app:diapara}
Next we study the zero-field orbital susceptibility defined as $\chi_0=L^{-1}(\partial J/\partial\phi)|_{\phi=0}$.
For $\chi_0>0$ ($\chi_0<0$) we speak of paramagnetic (diamagnetic) behavior of the orbital current is induced by the
gauge field. At low temperatures where only the lower band is occupied, we obtain:
\begin{align}
\chi_0&=-\frac{2}{2\pi}\int_{0}^{k^0_F} \frac{t}{2}\left[ \cos(q)-\frac{4t\sin^2(q)}{\Omega} \right] dq\notag\\
&=-\frac{t}{2\pi}\left[ \sin(n\pi)+\frac{t(\sin(2n\pi)-2n\pi)}{\Omega}\right].\label{eq:chi0}
\end{align}
In the last line, we have used that $n=k^0_F/\pi$. The boundary for the transition from paramagnetic to 
diamagnetic behavior is obtained by solving the following equation:
\begin{equation}
\frac{t}{2\pi}\left[ \sin(n\pi)+\frac{t(\sin(2n\pi)-2n\pi)}{\Omega}\right]=0,\label{eq:chi00}
\end{equation}
for the hopping strength ratio $\Omega/t$ and the particle density $n$.
Fig.~\ref{fig:j0} shows this boundary between the diamagnetic and paramagnetic behavior
obtained in this fashion.


\section{Orbital susceptibility near Lifshitz transitions}~\label{app:constraint}

In this appendix, we obtain exact expressions for the energy density, orbital current and susceptibility in the thermodynamic for the non-interacting system at zero temperature. We consider the two different types of constraints: fixed chemical potential and fixed particle number.
\subsection{Thermodynamic limit}
In the large $\Omega$ limit of interest in this work and at zero temperature, it is safe to neglect the contribution form the upper band. Thus, the following expression for energy density in thermodynamic limit is obtained.
\begin{align}
\epsilon(\phi)&= \int_\text{1BZ} \frac{dq}{2\pi} \, n_{q,d} \epsilon_{d}(q,\phi) =\frac{2}{2\pi}\int_{k_{F1(\phi)}}^{k_{F2(\phi)}} \epsilon_{d}(q,\phi) dq,\\
&=\frac{1}{\pi}\left\{\tilde{\epsilon}\left[\phi,k_{F2}(\phi)\right]-\tilde{\epsilon}\left[\phi,k_{F1}(\phi)\right] \right\},\\
&=\frac{1}{\pi}\tilde{\epsilon}(\phi,k_{F}(\phi))\big|_{1}^2\label{eq:e}
\end{align}
where $0 < k_{F1}(\phi) < k_{F2}(\phi)$ are the Fermi momenta in the four Fermi-point phase. In the case of the two 
Fermi-point phase, we should take $k_{F1}(\phi)=0$. Finally, for the insulator phase for which the number of Fermi points is zero,  
$ k_{F1}(\phi)=0, k_{F2}(\phi)=\pi$. The values of the Fermi momenta are obtained below  for both fixed chemical potential and fixed particle number. In the above expression, we have also introduced the notation $f(k_F)|_1^2=f(k_{F2})-f(k_{F1})$ for a given function $f(k_F)$. A closed form for the function $\tilde{\epsilon}(\phi,k_F)$ in Eq.~\eqref{eq:e}  can be given  in terms of the incomplete elliptic integral of the second kind:
\begin{align}
E(k,m)=\int_{0}^{k}\sqrt{1-m\sin^2\theta}d\theta.
\end{align}
Hence, the energy density can be obtained from
\begin{align}
\tilde{\epsilon}(\phi,k_F)&=-2t\cos(\phi/2)\sin(k_F)-\frac{\Omega}{2}E(k_F,-g^2),
\end{align}
where $g=4t\sin(\phi/2)/\Omega$. The orbital current density at fixed chemical potential,
\begin{equation}
J_{\mu}=\frac{-1}{L\beta}\frac{\partial G(\beta,\mu,\phi)}{\partial\phi}  =\frac{1}{L} \big\langle \left( \frac{\partial H}{\partial \phi}-\mu
\frac{\partial N}{\partial\phi} \right) \big\rangle,
\end{equation}
where $G(\beta,\mu,\phi)$ is the thermodynamic potential. For fixed number of particles, it is 
\begin{equation}
 J_n=\frac{-1}{\beta} \frac{\partial}{\partial\phi}\left[G(\beta,\mu,\phi)/L+n\mu\right]  
 =\frac{1}{L} \Big\langle\frac{\partial H}{\partial \phi}\Big\rangle.
 \end{equation}
As explained in Sec.~\ref{sec:intro},  
at zero temperature , the orbital current with fixed particle number and chemical potential are given by
\begin{align}
J_{n}(\phi)&=\frac{1}{\pi}\frac{
\partial \tilde{\epsilon}(\phi,k_F(\phi))}{\partial \phi}\Big|_{1}^{2}=\frac{1}{\pi}\tilde{J}(\phi,k_F(\phi))\Big|_{1}^{2},\\
J_{\mu}(\phi)&=J_{n}(\phi)- \mu \left(\frac{\partial n(\phi)}{\partial \phi} \right) \label{eq:Ju}.
\end{align}
From these expressions and in the limit of large $\Omega$, 
we can derive the  orbital current density using 
the following relations for the elliptic integral function:
\begin{align}
&\frac{\partial E(k,m)}{\partial k}=\sqrt{1-m\sin^2(k)},\\
&\frac{\partial E(k,m)}{\partial m}=\frac{E(k,m)-F(k,m)}{2m},
\end{align}
where $F(k,m)$ is the incomplete elliptic integral of the first kind:
 \begin{align}
F(k,m)=\int_{0}^{k}\frac{1}{\sqrt{1-m\sin^2\theta}}d\theta.
\end{align}
Hence,
\begin{align}
\tilde{J}(\phi,k_F)&=\frac{1}{\pi}\frac{\partial k_F}{\partial \phi} \epsilon(\phi,k_F)
+t\sin(\phi/2)\sin(k_F)\notag\\
&-\Omega\cot(\phi/2)\frac{E(k_F,-g^2)-F(k_F,-g^2)}{4}\label{eq:J}
\end{align}
We note that, for fixed chemical potential, the first term on right hand-side of the above expressionn  is exactly 
cancelled by the second term in Eq.~\eqref{eq:Ju} since $\epsilon(k_F,\phi)=\mu$. 

 The susceptibility is obtained from the derivative of orbital current using the following  identities:
\begin{align}
\frac{\partial F(k,m)}{\partial k}&=\frac{1}{\sqrt{1-m\sin^2(k)}},\\
\frac{\partial F(k,m)}{\partial m}&=-\frac{E(k,m)+(m-1)F(k,m)}{2m(m-1)},\notag
\\&+\frac{\sin(2k)}{4(m-1)\sqrt{1-m\sin^2(k)}}
\end{align}
We can express susceptibility by applying 
the derivatives to $\tilde{\epsilon}$, which yields:
\begin{align}
\chi_n(\phi)&=\frac{1}{\pi}\tilde{\chi}(\phi,k_F)\big|_{1}^{2},\\
\chi_{\mu}(\phi)&=\frac{1}{\pi}\tilde{\chi}(\phi,k_F)\big|_{1}^{2}-\frac{\partial^2 n(\phi)}{\partial\phi^2}\mu\label{eq:xmu}
\end{align}
where we have introduced the function
\begin{align}\label{eq:xx}
\tilde{\chi}(\phi,k_F)=\biggl[\frac{\partial^2 }{\partial\phi^2} +2\frac{\partial k_F}{\partial\phi}  & \frac{\partial }{\partial k_F\partial \phi}+\frac{\partial^2 k_F}{\partial\phi^2}\frac{\partial}{\partial k_F}\notag\\
&+\left(\frac{\partial k_F}{\partial\phi}\right)^2\frac{\partial }{\partial k_F^2}\biggr]\tilde{\epsilon}(\phi,k_F),\notag\\
=\frac{1}{\pi}\biggl[2\frac{\partial k_F}{\partial\phi}  \frac{\partial }{\partial \phi}+\frac{\partial^2 k_F}{\partial\phi^2}&+\left(\frac{\partial k_F}{\partial\phi}\right)^2 \frac{\partial }{\partial k_F} \biggr]\epsilon(\phi,k_F)\notag\\
&+\frac{1}{\pi}\frac{\partial^2 \tilde{\epsilon}(\phi,k_F)}{\partial\phi^2}.
\end{align}
In the second equation we have used  $\partial _{k_F} \tilde{\epsilon}(\phi,k_F)=\epsilon_d(\phi,k_F)$, i.e.
the dispersion relation of the lower band. Note that the last term involves a partial derivative with respect to $\phi$ 
and it can be also written as:
\begin{equation}
\frac{1}{\pi}\frac{\partial^2 \tilde{\epsilon}(\phi,k_F)}{\partial\phi^2} = \int^{k_F}_0 \frac{dq}{\pi}\,  \frac{\partial^2 \epsilon_d(q,\phi)}{\partial \phi^2}.
\end{equation}
The above function upon setting $k_F = k_{F1,2}(\phi)$ is a continuous function
of $\phi$. In the following section, we shall  also see that the Fermi momenta $k_{F1,2}(\phi)$
is continuous with $\phi$, but their derivatives are not. Thus, 
any possible singularities of the orbital susceptibility must arise from the terms
in Eq.~\eqref{eq:xx} that contain derivatives of the Fermi momenta with respect to $\phi$.
In connection to this, it is important to note that, in the case of fixed particle density, the susceptibility depends 
on the first and second order derivatives of the Fermi momenta with respect to $\phi$. However, for 
fixed chemical potential,  using Eqs.~\eqref{eq:xmu} and \eqref{eq:xx},  the second order derivative of  
$k_F$ with respect to $\phi$ is cancelled and the orbital susceptibility only depends on the first derivative.

\subsection{Numerical determination  of the Fermi points}

\begin{figure}[b]
 \center
\includegraphics[width=\columnwidth]{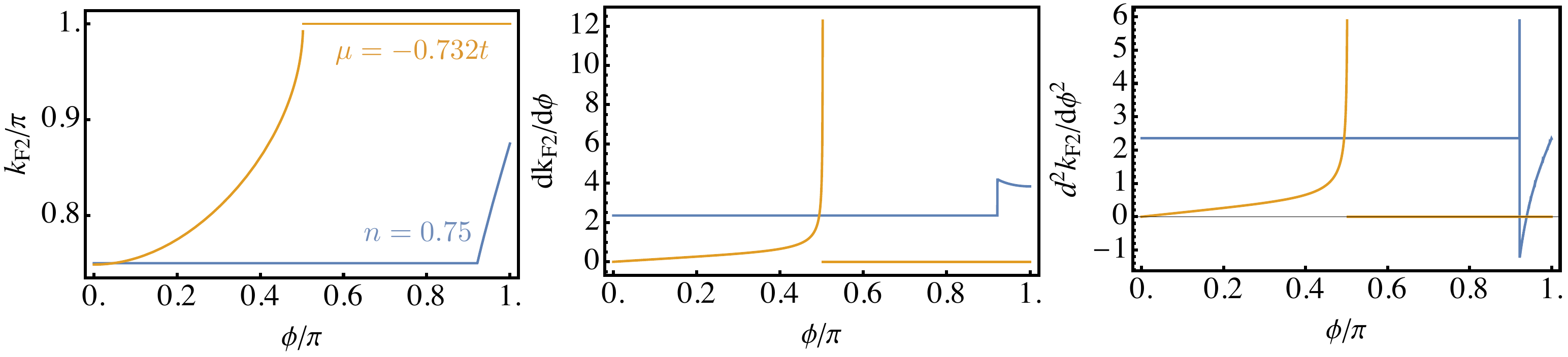}
\caption{The largest Fermi momentum, $k_{F2}$, and its derivatives as a function of $\phi$  for the types of constraints (fixed  particle density $n$ and fixed chemical potential $\mu$). We note that, for fixed $\mu$, the Fermi points shows a square root dependence which gives rise to a  $(\phi-\phi_c)^{-1/2}$ behavior  ($\propto \partial k_F/\partial \phi$) van Hove singularity. The second derivative term $\propto \partial ^2k_F/\partial \phi^2$) is cancelled(c.f. eq.\eqref{eq:xx} and the following context).  For fixed particle number, the behavior gives a delta-function($\propto \partial^2 k_F/\partial \phi^2$) susceptibility at the transition point with a step-function ($\propto \partial k_F/\partial \phi$) behavior near the transition point.\label{fig:kf}}
 \end{figure}

For fixed chemical potential, the location Fermi momenta can be obtained from
\begin{align}
\epsilon_{d}(q,\phi)=\mu, 
\end{align}
which can be solved analytically.  For fixed particle number, the Fermi points are related by the 
relation:
\begin{equation}
n = \frac{N}{L} = \int^{k_{F2}(\phi)}_{k_{F1}(\phi)} \frac{dq}{\pi} = k_{F2}(\phi) - k_{F1}(\phi).
\end{equation}
This relation determines only one of the Fermi points. The other can be obtained by requiring that
\begin{align}
\epsilon_{d}(k_{F1}(\phi),\phi)&=\epsilon_{d}(k_{F2}(\phi),\phi) \\
&= \epsilon_{d}(k_{F1}(\phi) + \pi n,\phi). \label{eq:eqdis}
\end{align}
Note that in the two Fermi-point phase, we must set $k_{F1}=0$. In the four Fermi-point phase where $k_{1F}, k_{2F} \neq 0$,  the above equation can be solved numerically to high precision.

 Fig.~\ref{fig:kf} shows the behavior of  the (largest) Fermi momentum numerically obtained for $n=0.75$ and fixed chemical potential ($\mu=-0.372t$). On the same figure, we also show the first and second derivatives of the Fermi momentum.  For fixed chemical potential, the first derivative exhibits square root dependence with the magnetic flux. However, for fixed particle number, the first order derivative is a step function whereas the second order derivative contains both a step discontinuity and a Dirac delta-function  singularity (resulting from the discontinuity of the first derivative) at the critical point~\footnote{This result can be obtained analytically if we solve Eq.~\eqref{eq:eqdis} using a `toy model' dispersion $\epsilon(k,\phi)  = a(\phi) -   (\phi-\phi_0) k^2/2 m +   \lambda k^4/4$, with $a(\phi), m, \lambda > 0$. This dispersion has the same ``double well'' structure as the lower band dispersion
 close for $ \phi > \phi_0$. }
 Using these results  along with Eq.~\eqref{eq:xx}, we will next investigate the singularities of the orbital susceptibility 
near the critical point.

\subsection{Singular behavior of the orbital susceptibility}
\begin{figure}[t]
\includegraphics[width=\columnwidth]{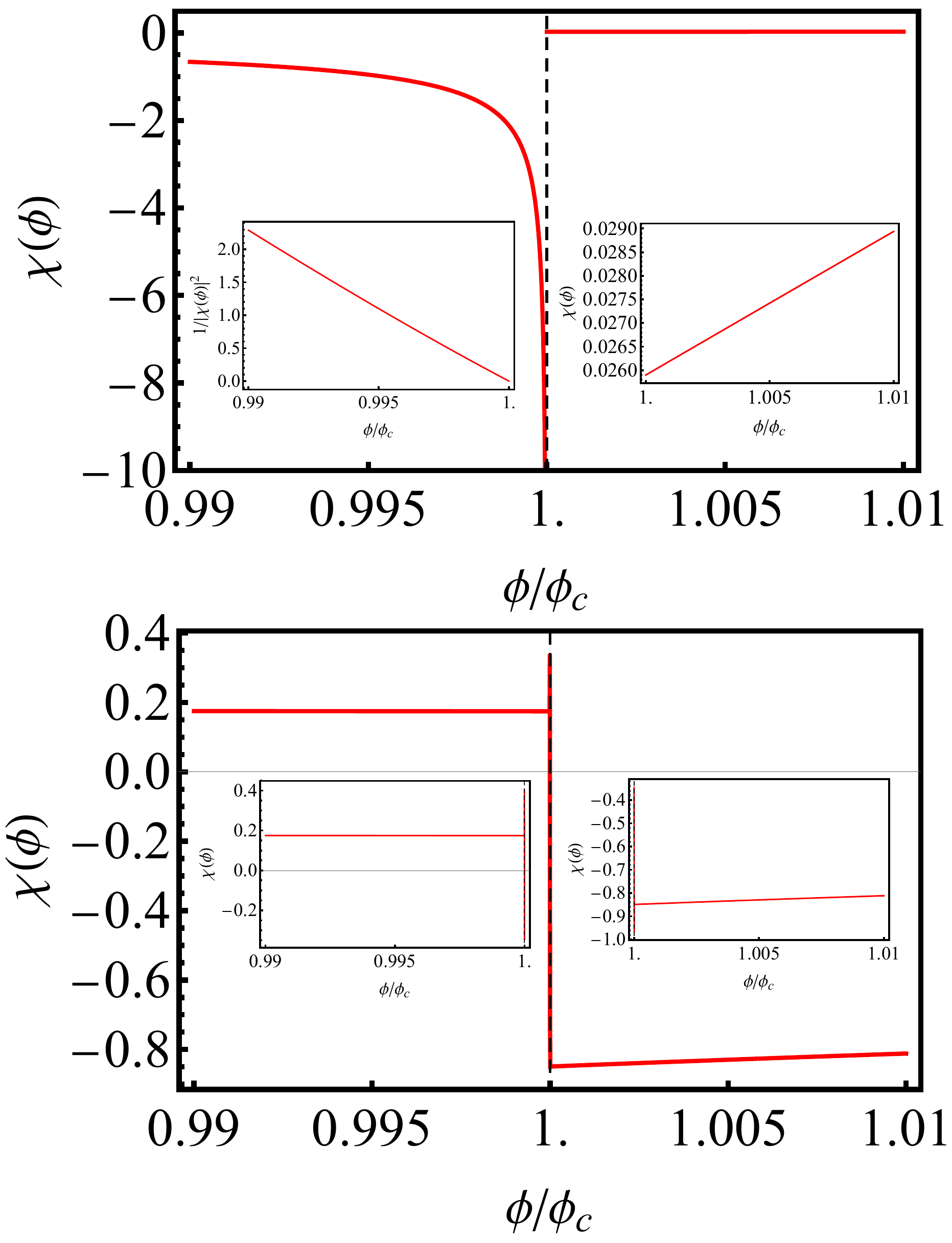}

\caption{ Orbital susceptibility near the critical point of the Lifshitz transition point for fixed chemical. \textbf{Upper panel:} For $\mu=-0.732t$, the susceptibility shows a $(\phi-\phi_c)^{-1/2}$  singularity when  the number of Fermi points changes from two (left) to zero (right). \textbf{Bottom panel:}  $|\chi(\phi)|$ for susceptibility near the singularity for fixed particle number ($n=0.75$). The singularity shows a step function near the Lifshitz transition point when the system changes number of Fermi points from two (left) to  four (right).}
\label{fig:xfix0}
\end{figure}

\begin{figure}[ht]
\includegraphics[width=\columnwidth]{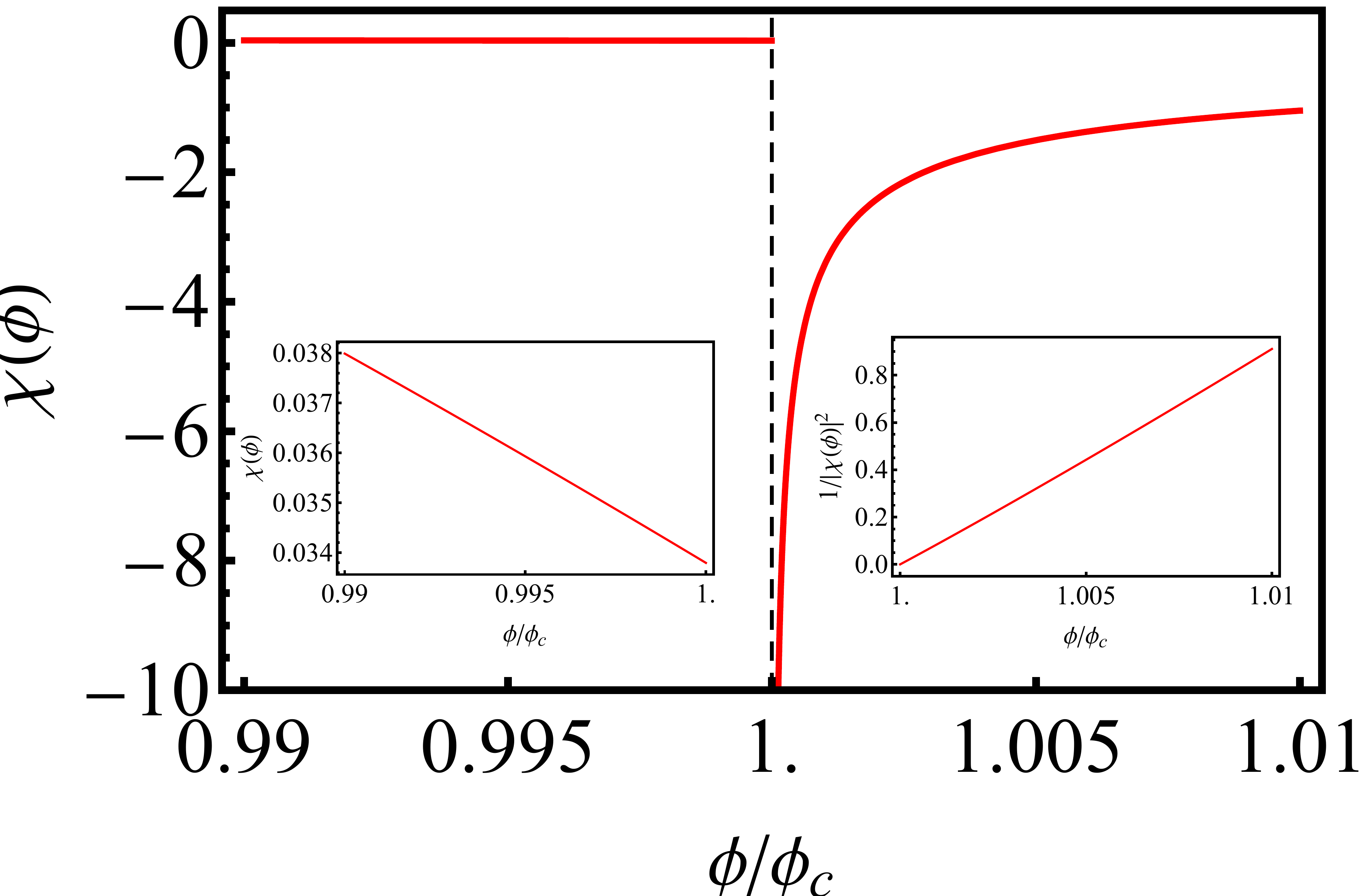}
\caption{ For $\mu=-2.5t$, the susceptibility shows a $(\phi-\phi_c)^{-1/2}$ singularity when the system 
undergoes a Lifshitz transition where the  Fermi points changes from two (right) to four (left).}
\label{fig:xfix1}
\end{figure}

 Using the previous results,  below we obtain the critical behavior of the orbital susceptibility for fixed
 chemical potential and fixed particle number.   Plots of the   susceptibility for fixed particle number and fixed chemical potential is shown in Figs.~\ref{fig:jmu} and \ref{fig:xfix0}. The results of Fig.~\ref{fig:jmu} are obtained for a finite systems for which the singularities in the orbital current and susceptibility  are slightly broadened by the finite-size effects. To investigate the singular behavior, we have plotted in Fig.~\ref{fig:xfix0} the same quantity in the thermodynamic limit as obtained from the above formulas and focusing on the regime near the critical point of the  Lifshitz transition. For fixed $\mu=-0.732t$ (lower panel of Fig. \ref{fig:xfix0}), an inverse square-root divergence   $\sim (\phi-\phi_c)^{-1/2}$ is observed. In this case, 
 the system undergoes a transition between a metal with two Fermi points and a  band insulator. The same critical behavior is observed at another choice of fixed $\mu=-2.5t$, which is shown on Fig.~\ref{fig:xfix1}. In the latter case,  the Lifshitz transition takes place by  changing the number of Fermi points changing between two and four. These results demonstrate that the singular behavior at the critical point is entirely determined by the constraint of fixed chemical potential and not by the type of Lifshitz transition. The inverse square-root singularity  arises from the 
van Hove singularity in the density of states at the top of the lower band because, as $\phi$ is varied, the top of the lower band must 
cross the Fermi level. Using the results obtained in the previous subsection, we can see that the inverse square-root singularity stems from
the same singularity displayed by $\partial_{\phi} k_F(\phi)$ (cf.  Fig.~\ref{fig:kf}).

 Finally, let us consider the case of fixed particle number.  Using Eq.~\eqref{eq:xx} shows that $\chi_n$ we see
 that $\chi_n$ depends both of the first and second derivative of the Fermi momentum. For this constraint,  
 the Fermi momentum close to the Lifshitz transition $k_{F2}(\phi)$ grows from a 
 constant value approximately linearly (cf. Fig.~\ref{fig:kf}). $k_{2F}$ ($k_{1F}$) is continuous, but its first and second derivatives, 
 upon which $\chi_n$ depends, exhibit both step-like discontinuities and a Dirac-delta
 singularity, which can be also seen on Fig.~\ref{fig:xfix0}  (lower panel) and Fig.~\ref{fig:jmu}  (right panel) for $\chi_n$.

\section{Mean-field theory}~\label{app:MF}



\begin{figure}[t]
\includegraphics[width=\columnwidth]{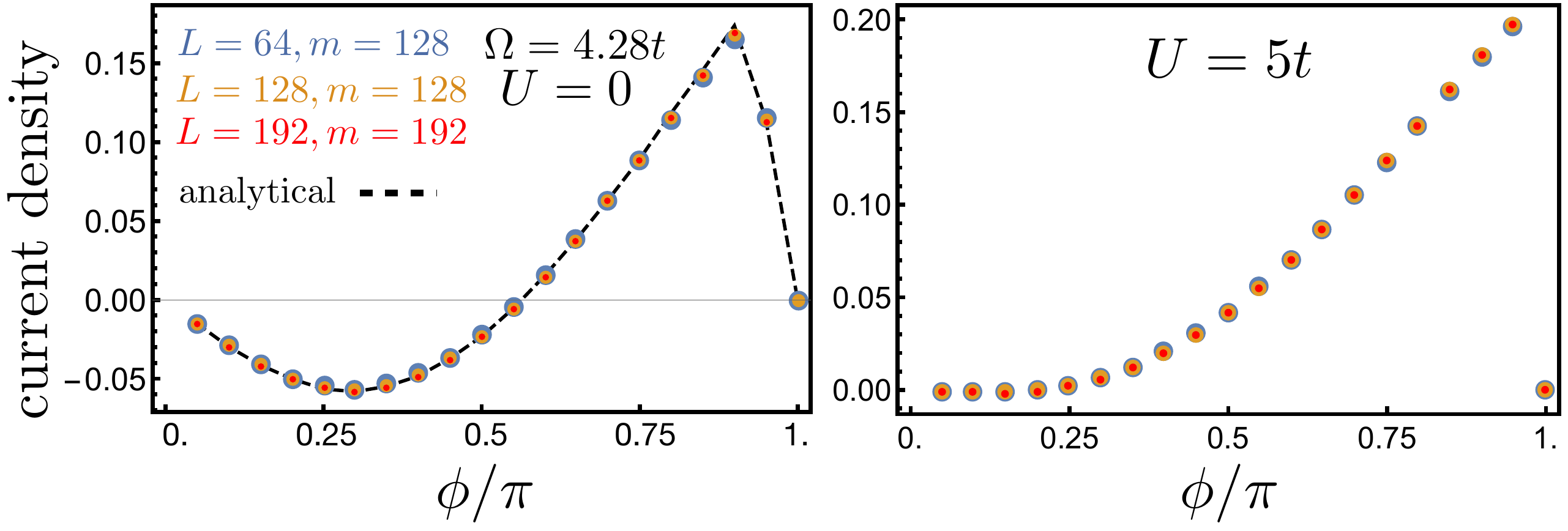}
\caption{ Convergence of the DMRG result for $\Omega=4.28,n=0.75,U=0$ and $U=5t$ for different system length $L$, and the number of block states $m$.  The choice of $L=128$ and $m=128$ gives deviation from the analytical result less than $1\%$.}\label{fig:conv}
\end{figure}
%


 The mean-field Hamiltonian in Eq.~\eqref{eq:MF} can be solved numerically on a finite lattice by exact diagonalization, which allows us to self-consistently determine the mean-field parameters. Let us introduce a vector notation where $\mathbf{u}^{\dag}$ is given by
\begin{align}
\mathbf{u^{\dag}}=\left\{c^{\dag}_{\uparrow,1},c^{\dag}_{\downarrow,1},c^{\dag}_{\uparrow,2},c^{\dag}_{\downarrow,2},...,c^{\dag}_{\uparrow,L},c^{\dag}_{\downarrow,L},\right\}.
\end{align}
Thus, the mean-field Hamiltonian can be written as follows:
\begin{align}
\hat{H}^{\text{MF}}=\mathbf{u^{\dag}}\mathcal{H}^{\text{MF}}\mathbf{u}.
\end{align}
Let $\{|\mathbf{v_i}\rangle=\left\vert\nu^{i}_{1\uparrow} ,\nu^{i}_{1\downarrow} ,...,\nu^{i}_{L\uparrow},\nu^{i}_{L\downarrow} \right\rangle\}$ 
be the eigenvectors of the mean-field Hamiltonian matrix
$\mathcal{H}^{\text{MF}}$  with eigenvalues $\{\epsilon_i\}$, i.e.
\begin{align}
\mathcal{H}^{\text{MF}}|\mathbf{v_i}\rangle=\epsilon_i |\mathbf{v_i}\rangle.
\end{align}
Hence, the mean-field parameters can be obtained from the following expressions:
\begin{align}
\alpha_m(\phi) &= \sum_{<i>} (\nu_{m\uparrow}^{i})^* \nu_{m\downarrow}^i,\\
n_{m,\uparrow}&= \sum_{<i>} (\nu_{m\uparrow}^{i})^*\nu_{m\uparrow}^i,\\
n_{m,\downarrow} &= \sum_{<i>} (\nu_{m\downarrow}^{i})^* \nu_{m\downarrow}^i.
\end{align}
The latter are a set of equations that  need to be solved self-consistently
 since the coefficients $\nu_{m\sigma}^i$ depend on the mean-field parameters themselves. Note that in the above equation we have used the notation    $\sum_{\left\langle i\right\rangle}$, which means that the summation of the lowest $N=L\times n$ eigenstates for $N$ particles at zero temperature. 
 From the self-consistent solution, e.g. the total orbital current on the $\sigma=\uparrow$-leg can be obtained as follows:
\begin{align}
J^{\text{MF}}_{\uparrow}=2\, \mathrm{Im}\left[\sum_m\sum_{\left\langle i\right\rangle} \left( \nu_{m,\uparrow}^{i}
\right)^*\nu_{m+1,\uparrow}^i e^{i \phi/2}\right]=\sum_{m}j_{m,\uparrow},
\end{align}
where $j_{m,\uparrow} $ is the local orbital current.


\section{Convergence of the DMRG results}~\label{app:DMRG}

Fig.~\ref{fig:conv} shows the DMRG results of unperturbed current with different system sizes and block states and the analytical result obtained in thermal dynamic limit. The weak dependence on the system length $L$ and the number of block states indicates that our numerical results can be trusted for discussing the thermodynamic limit of the system.

\bibliography{2leg}
\end{document}